\documentclass[twocolumn,prb,citeautoscript,amsmath,amssymb,floatfix,eqsecnum]{revtex4}
\usepackage[utf8]{inputenc}
\usepackage{amsmath}
\usepackage{amsfonts}
\usepackage{amssymb}
\usepackage{graphicx}
\usepackage{hyperref}
\usepackage{bm}

\newcommand{\kv}{{\bm{k}}}
\newcommand{\qv}{{\bm{q}}}
\newcommand{\qkv}{{{\bm{q}}-{\bm{k}}}}
\newcommand{\kqv}{{{\bm{k}}-{\bm{q}}}}
\newcommand{\vfv}[1]{\tilde{\psi}_{#1}}
\newcommand{\vfe}[1]{\tilde{\psi}_{#1}^*}

\newcommand{\gam}[3]{\Gamma_{#1;#2,#3}^{\bar{a}aa}}
\newcommand{\pvf}[1]{\tilde{p}_{#1}^c}
\newcommand{\bd}{\bm}

\begin{document}
\title{Effect of magnon decays on parametrically pumped magnons}
\date{\today}
\author{Viktor Hahn}
\email{hahn@itp.uni-frankfurt.de}
\author{Peter Kopietz}
\affiliation{Institut f\"{u}r Theoretische Physik, Universit\"{a}t Frankfurt, Max-von-Laue Strasse 1, 60438 Frankfurt, Germany}
\begin{abstract}
We investigate the influence of magnon decays on the non-equilibrium 
dynamics of parametrically excited magnons in the magnetic insulator yttrium-iron garnet (YIG). 
Our investigations are motivated by a recent experiment by Noack 
{\it{et al.}} [Phys. Status Solidi B {\bf{256}}, 1900121 (2019)]
 where an enhancement of the spin pumping effect in YIG was observed near 
the magnetic field strength where magnon decays via confluence of magnons 
becomes kinematically possible.
To explain the experimental findings, 
we have derived and solved  kinetic equations for the
non-equilibrium magnon distribution. The effect of magnon decays 
is taken into account microscopically via collision integrals derived from
interaction vertices involving three powers of magnon operators.
Our results agree quantitatively with the experimental data.
\end{abstract}
\maketitle

\section{Introduction}

In a recent experiment \cite{Noack19} the parametric excitation of magnons
in the magnetic insulator yttrium-iron garnet (YIG) was investigated by
coupling an oscillating microwave field into the system and measuring the
magnon density via the inverse spin-Hall effect\cite{Hirsch99}.
This effect, which
converts a spin current into an electric field perpendicular to the
directions of the spin current and the 
spin polarization,  is caused by the relativistic spin-orbit interactions
that are also responsible for the direct spin-Hall effect\cite{Ando11}.
In solids this effect is enhanced due to the strong potential of atomic nuclei \cite{Nagaosa08}. In the experiment\cite{Noack19} a thin YIG film 
was  exposed to an oscillating magnetic field
$ \bd{H} ( t ) =  {H}_0 \bd{e}_z +  H_1 \cos ( \omega_0 t ) 
 \bd{e}_z$, where the static part $H_0 \bd{e}_z$ forces
the macroscopic magnetization to be aligned along the $z$-axis $\bd{e}_z$,
while  the oscillating part with amplitude
$H_1 \ll H_0$ drives the magnons in the sample out of equilibrium.
Noack {\it{et al.}} \cite{Noack19} observed that 
the spin-pumping effect was enhanced for certain values of the 
static field $H_0$, and that
the magnon density in the stationary non-equilibrium state displayed 
peaks or dips for those values of $H_0$ where magnon  
decays due to the confluence of two parametrically excited magnons 
with identical energy and momentum becomes kinematically possible. Recall that 
magnon decays due to  the 
confluence and the reverse splitting process conserve 
the total energy and momentum of the magnons 
involved in these scattering processes~\cite{Cherepanov93,Lvov94}.

In this work we  provide a quantitative microscopic
explanation for the experimental observations of Ref.~[\onlinecite{Noack19}].
It turns out that therefore a proper understanding of magnon damping
under non-equilibrium conditions in YIG
is crucial. We therefore construct a kinetic theory of pumped magnon gases
including microscopically derived collision integrals describing 
the relevant dissipative effects. 
While theoretical investigations of pumped magnon gases in magnetic insulators
have a long history \cite{Suhl57,Schloemann60,Zakharov70,Zakharov74,Vinikovetskii79,Cherepanov93,Araujo74,Tsukernik75,Lavrinenko81,Zvyagin82,Zvyagin85,Lim88,Kalafati89,Lvov94,Zvyagin07,Rezende09,Kloss10,Safonov13,Slobodianiuk17,Hahn20}
in all works published so far
the effect of collisions on the non-equilibrium magnon dynamics was 
considered only phenomenologically by introducing (by hand) 
a relaxation rate into the kinetic equations for the magnon distribution functions. Although the relevant microscopic collision integrals 
have been derived within the Born approximation in Ref.~[\onlinecite{Vinikovetskii79}], to our knowledge a microscopic treatment of the effect of magnon collisions 
on the non-equilibrium dynamics of magnons is still missing in the literature.
An alternative method to investigate the dynamics of pumped magnons in YIG
is based on the  numerical solution of the 
stochastic non-Markovian Landau-Lifshitz-Gilbert equation 
with a microscopically derived noise and dissipation kernel \cite{Rueckriegel15}. 
The approach based on kinetic equations adopted here has the advantage that it
allows us to identify the experimentally relevant confluent scattering processes 
directly in the collision integral.
Still,  the resulting non-linear integro-differential equations are very complicated and can only be solved numerically. Moreover, the derivation of the
collision integrals starting from an effective spin Hamiltonian for YIG
is a demanding technical problem because the distribution function
of the magnon gas in YIG 
with external pumping has an off-diagonal  component 
so that we have to deal with various types of anomalous
cubic interaction vertices. 
While in principle the collision integrals can be derived diagrammatically using the Keldysh formalism\cite{Kamenev11},
to keep track of all terms contributing to the collision integrals
we have found  it more convenient to use an unconventional method 
developed in Ref.~[\onlinecite{Fricke97}]
based on a systematic expansion of the collision integrals in terms of connected equal-time correlation functions.

The rest of this 
article is organized as follows. 
In Sec.~\ref{sec:eff_H} we introduce the effective Hamiltonian 
describing pumped magnons in YIG
which is the starting point for our investigations. In Sec.~\ref{sec:kinetic_eq} we derive 
collisionless kinetic equations for the magnon distribution functions in YIG. 
We also discuss the usual  phenomenological strategy of 
introducing  dissipative effects into 
the collisionless kinetic equations, derive the resulting
stationary non-equilibrium distributions for YIG, and show that the experimental results 
of Noack {\it{et al.}} \cite{Noack19} cannot be explained within this approximation. 
In Sec.~\ref{sec:col_int_der} we  derive the collision integrals containing the cubic vertices
using an expansion in powers of connected equal-time correlations \cite{Fricke97}.
Our numerical results for the stationary non-equilibrium solution including the effects of the cubic vertices are presented in Sec.~\ref{sec:md_cv}. Finally, in Sec.~\ref{sec:conclusions} we summarize our results and present our conclusions.
To make this work self-contained we have added three appendices with technical details.
In Appendix~A we outline the derivation of the Hamiltonian of 
pumped magnons in YIG following mainly Refs.~[\onlinecite{Kreisel09,Hick10}].
In Appendix~B we review the method of deriving kinetic equations via an expansion in terms of connected equal-time correlations developed by Fricke \cite{Fricke97}, and in Appendix~C
we give the explicit expressions for the relevant collision integrals for YIG 
obtained with this method.

\section{\label{sec:eff_H}
Hamiltonian for pumped magnons in YIG}

\begin{figure}
 \centering
 \includegraphics[width=0.8\linewidth,clip=true,trim=30pt 25pt 45pt 55pt]{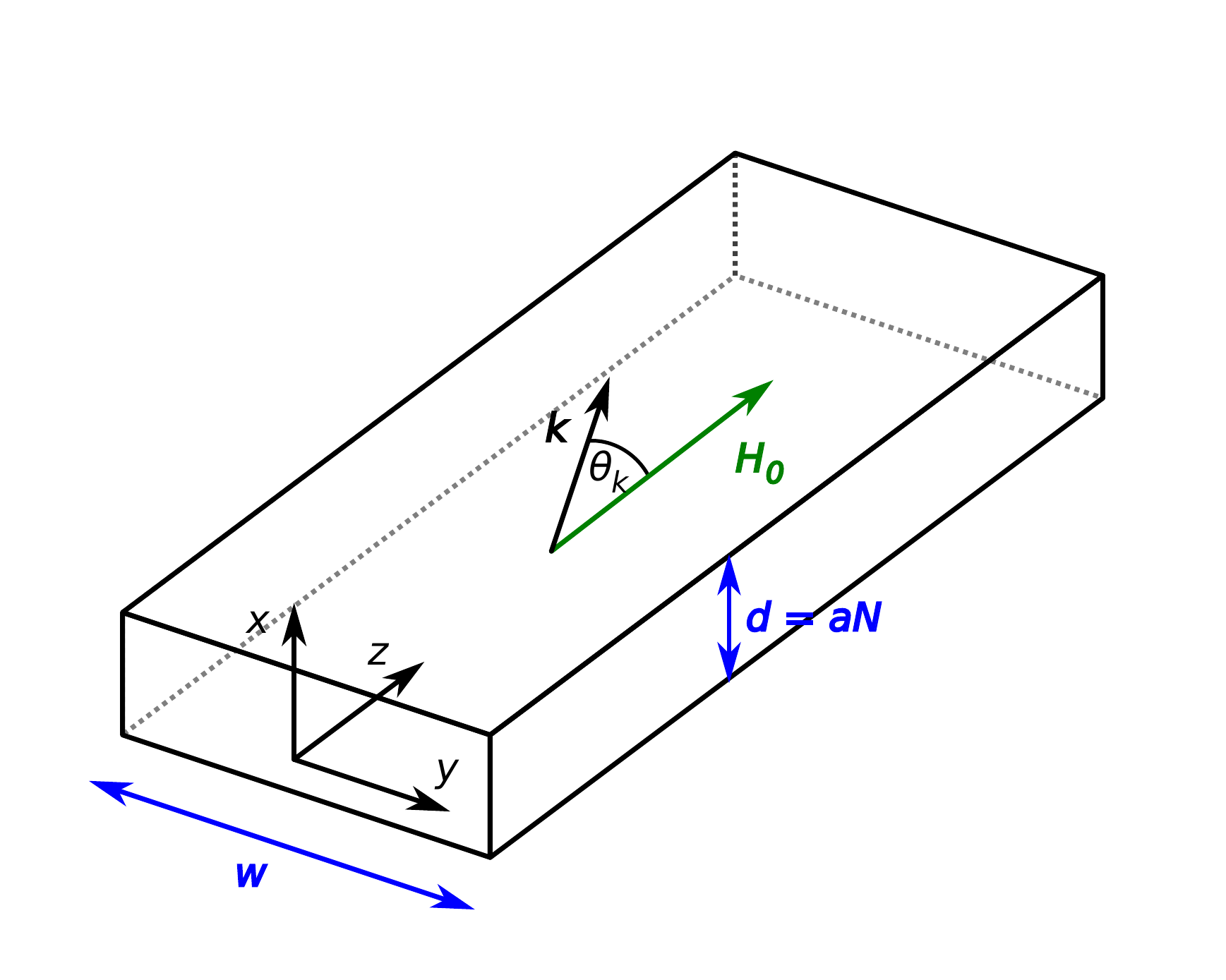}
 \caption{Sketch of a long YIG stripe 
oriented along the $z$-axis 
 with width $w$ in $y$-direction and thickness $d=aN$ in $x$-direction.
Here $a$ is the lattice spacing and $N$ is the number of lattice sites in
$x$-direction.
In this work we consider wavevectors $\kv$ in the $y$-$z$-plane with $\theta_\kv$ being the angle between $\kv$ and the  static magnetic field  magnetic field ${\bd H}_0 = H_0 {\bf e}_z$.
}
 \label{fig:setup}
\end{figure}

In the experimental setup of Ref.~[\onlinecite{Noack19}]
a thin stripe of YIG
is exposed to an oscillating microwave field in the parallel pumping
geometry where the oscillating component of the magnetic field is parallel to
its static component.
At the energy scales of interest the magnon dynamics can be described by the following
time-dependent effective Hamiltonian,\cite{Cherepanov93,Lvov94,Rezende06,Rezende09,Kreisel09,Hick10,Kloss10,Rueckriegel14}
\begin{eqnarray}
{\mathcal{H}}(t)&=&-\frac{1}{2}\sum\limits_{i j}\sum\limits_{\alpha \beta}\left[J_{i j}\delta^{\alpha \beta} + D_{i j}^{\alpha \beta}\right]S_i^\alpha S_j^\beta \nonumber\\*
&&-\left[h_0+h_1\cos\left(\omega_0 t\right)\right]\sum\limits_i S_i^z,
\label{eq:H_spin}
\end{eqnarray}
where the  indices $i,j $ label the  $N$ sites of a cubic lattice 
and $\alpha,\beta$ denote  the three spin components $x,y,z$ of the spin operators $S_i^\alpha$. The nearest neighbor exchange couplings
connecting lattice sites $\bd{r}_i$ and $\bd{r}_j$ are denoted by $J_{ij}$,
while  $D_{ij}^{\alpha\beta}$ denotes the matrix elements of
the dipolar tensor defined in Eq.~(\ref{eq:Ddef}) of Appendix~A. 
The last term in Eq.~(\ref{eq:H_spin})
represents the coupling of the spins 
to a static magnetic field $H_0$ and a time-dependent
microwave magnetic field $H_1$ oscillating with frequency $\omega_0$, where $h_0=\mu H_0$ and $h_1=\mu H_1$ are the corresponding
Zeeman energies. The geometry of the system and our choice of the
coordinate system is shown in Fig.~\ref{fig:setup}.
The Hamiltonian \eqref{eq:H_spin} can be bosonized  
using the Holstein-Primakoff transformation \cite{Holstein40} as described in Appendix~A.
We expand the resulting bosonized  Hamiltonian 
in powers of the inverse spin quantum number $1/S$,
\begin{equation}
\mathcal{H}(t)=\mathcal{H}_0(t)+\mathcal{H}_2(t)+\mathcal{H}_3+\mathcal{H}_4+\mathcal{O}(S^{-1/2}),
\label{eq:H_1/S}
\end{equation}
where $\mathcal{H}_n$ contains $n$ powers of the boson operators.
Explicit expressions for the terms in the expansion (\ref{eq:H_1/S}) 
are given in Refs.~[\onlinecite{Kreisel09,Kloss10,Hick10,Hahn20}] and are 
reproduced in Appendix~A.
It is convenient to use a canonical (Bogoliubov) transformation to
diagonalize the time-independent part of ${\cal{H}}_2 ( t )$, which then assumes
the form given in Eq.~(\ref{eq:H2_r}).
For our purpose it is sufficient to further simplify 
${\cal{H}}_2 ( t )$ by dropping all non-resonant terms which are explicitly 
time-dependent in the
rotating reference frame defined by the canonical transformation 
(\ref{eq:rotref}) below \cite{Kloss10,Hick10,Hahn20}. 
In this approximation
\begin{eqnarray}
\mathcal{H}_2\left(t\right) &=& \sum\limits_\kv \left[\epsilon_\kv a^\dagger_\kv a_\kv 
+ \frac{1}{2} V_\kv \text{e}^{-i \omega_0 t} a^\dagger_\kv a^\dagger_{-\kv} \right.\nonumber\\*
&& \hspace{19.5mm}\left. + \frac{1}{2} V_\kv^* \text{e}^{i \omega_0 t} a_{-\kv} a_\kv\right],
\label{eq:H_2}
\end{eqnarray}
where  $a_\kv$ and
$a_\kv^\dagger$ annihilate and create  magnons with momentum $\bd{k}$ and
energy $\epsilon_\kv$. For small $\bd{k}$  the magnon energy
can be approximated 
by \cite{Kreisel09,Tupitsyn08,Kalinkos86}
\begin{equation}
\epsilon_\kv=\sqrt{\left[h_0+\rho \kv^2+\left(1-f_\kv\right)\Delta\sin^2\theta_\kv\right]\left[h_0+\rho\kv^2+f_\kv\Delta\right]},
\label{eq:disp_rel}
\end{equation}
while the pumping energy $V_\kv$ can be written as
\begin{equation}
V_\kv=\frac{h_1\Delta}{4\epsilon_\kv}\left[-f_\kv+\left(1-f_\kv\right)\sin^2\theta_\kv\right].
\label{eq:pump_en}
\end{equation}
Here, $\rho$ is  the exchange stiffness of long-wavelength magnons,\cite{Kreisel09} 
the dipolar energy scale
 \begin{equation}
 \Delta = \frac{4 \pi \mu^2 S}{a^3}
 \label{eq:deltadipdef}
 \end{equation}
is determined by the effective magnetic moment $\mu$ and the effective spin $S$
[see Eq.~(\ref{eq:Deltadef})], and the form factor $f_{\bd{k}}$ for a thin stripe of YIG shown in Fig.~\ref{fig:setup}
 is given by \cite{Kalinkos86,Kreisel09}
\begin{equation}
f_\kv = \frac{1 - \text{e}^{-\left|\kv\right|d}}{\left|\kv\right|d},
\label{eq:formdef}
\end{equation}
where $d$ is the thickness of the YIG stripe.
We parametrize the in-plane wavevector  
as 
 \begin{equation}
\kv = k_y\bm{e}_y+k_z\bm{e}_z = |\kv|\left(\sin\theta_\kv\bm{e}_y+\cos\theta_\kv\bm{e}_z\right),
 \label{eq:inplanewave}
 \end{equation}
where $\theta_\kv$ is the angle between the wavevector $\kv$ and the static 
magnetic field $H_0 \bd{e}_z$ as shown in  Fig.~\ref{fig:setup}.

The explicit time-dependence of the quadratic part of the Hamiltonian \eqref{eq:H_2} can be removed via  a canonical transformation to the rotating reference frame,
\begin{eqnarray}
\tilde{a}_\kv &=& \text{e}^{i \frac{\omega_0}{2} t} a_\kv,   \; \; \; 
\tilde{a}^\dagger_\kv = \text{e}^{-i \frac{\omega_0}{2} t} a^\dagger_\kv.
\label{eq:rotref}
\end{eqnarray}
The quadratic part of the Hamiltonian then becomes~\cite{Kloss10,Hick10,Hahn20}
\begin{equation}
\tilde{\mathcal{H}}_2 = \sum\limits_\kv \left[E_\kv \tilde{a}_\kv^\dagger \tilde{a}_\kv + \frac{V_\kv}{2} \tilde{a}_\kv^\dagger \tilde{a}_{-\kv}^\dagger  + \frac{V_\kv^*}{2} \tilde{a}_{-\kv} \tilde{a}_\kv\right],
\label{eq:H2_t}
\end{equation}
where 
 \begin{equation} 
 E_\kv=\epsilon_\kv-\omega_0/2
 \end{equation}
 is the shifted magnon energy in the rotating reference frame. It turns out that in 
this frame the cubic and the quartic parts of the magnon Hamiltonian
acquire an explicit time-dependence. Explicitly, after Bogoliubov transformation and
transformation the cubic and quartic part of the magnon Hamiltonian are in the rotating
reference frame of the form
\begin{widetext}
\begin{eqnarray}
\tilde{\mathcal{H}}_3(t) &=& \frac{1}{\sqrt{N}} \sum\limits_{\kv_1, \kv_2, \kv_3} \delta_{\kv_1+\kv_2+\kv_3, 0} \left[\frac{1}{2} \Gamma^{\bar{a}aa}_{1; 2, 3} e^{-i \omega_0 t / 2} \tilde{a}^\dagger_{-1} \tilde{a}_2 \tilde{a}_3 + \frac{1}{2} \Gamma^{\bar{a}\bar{a}a}_{1, 2; 3} e^{i \omega_0 t / 2} \tilde{a}^\dagger_{-1} \tilde{a}^\dagger_{-2} \tilde{a}_3 \right.\nonumber\\*
&&\hspace{35mm}\left.+ \frac{1}{3!} \Gamma^{aaa}_{1, 2, 3} e^{-3 i \omega_0 t/2} \tilde{a}_1 \tilde{a}_2 \tilde{a}_3 + \frac{1}{3!} \Gamma^{\bar{a}\bar{a}\bar{a}}_{1, 2, 3} e^{3 i \omega_0 t/2} \tilde{a}^\dagger_{-1} \tilde{a}^\dagger_{-2} \tilde{a}^\dagger_{-3}\right],
\label{eq:H3rot}\\
\tilde{\mathcal{H}}_4 (t) &=& \frac{1}{N} \sum\limits_{\kv_1, \kv_2, \kv_3, \kv_4} \delta_{\kv_1+\kv_2+\kv_3+\kv_4, 0} \left[\frac{1}{\left(2!\right)^2} \Gamma^{\bar{a} \bar{a} a a}_{1, 2; 3, 4} 
\tilde{a}^\dagger_{-1} \tilde{a}^\dagger_{-2} \tilde{a}_3 \tilde{a}_4 + \frac{1}{3!} e^{ - i \omega_0 t} \Gamma^{\bar{a} a a a}_{1; 2, 3, 4} 
\tilde{a}^\dagger_{-1} \tilde{a}_2 \tilde{a}_3 \tilde{a}_4 
\right.\nonumber
\\*
&& \left. + \frac{1}{3!} e^{ i \omega_0 t} 
\Gamma^{\bar{a} \bar{a} \bar{a}a}_{1, 2, 3; 4} \tilde{a}^\dagger_{-1} \tilde{a}^\dagger_{-2} 
\tilde{a}^\dagger_{-3} \tilde{a}_4 + \frac{1}{4!} e^{ - 2 i \omega_0 t}
\Gamma^{aaaa}_{1, 2, 3, 4} \tilde{a}_1 \tilde{a}_2 \tilde{a}_3 \tilde{a}_4 + \frac{1}{4!} e^{ 2 i \omega_0 t} \Gamma^{\bar{a} \bar{a} \bar{a}\bar{a}}_{1, 2, 3, 4} 
\tilde{a}^\dagger_{-1} \tilde{a}^\dagger_{-2} \tilde{a}^\dagger_{-3} \tilde{a}^\dagger_{-4} \right] ,
\label{eq:H4rot}
\end{eqnarray}
\end{widetext}
where we have introduced the short notation 
$\kv_i \rightarrow i$ for the momentum labels. In 
Eqs.~\eqref{eq:cubic_vertices} and \eqref{eq:quartic_vertices} 
of Appendix~A we explicitly give the rather cumbersome expressions
for the vertices appearing in Eqs.~(\ref{eq:H3rot}) and (\ref{eq:H4rot}).
At the first sight it seems that
within the rotating-wave approximation we should  drop all oscillating terms
in Eqs.~(\ref{eq:H3rot}) and (\ref{eq:H4rot}).
However, as will be shown in Sec.~\ref{sec:col_int_der}, the collision integrals originating from the cubic part $\tilde{\mathcal{H}}_3 (t)$ of the Hamiltonian
contain products of two cubic vertices, so that some of the time-dependent factors in
Eq.~(\ref{eq:H3rot}) cancel  in the collision integrals and at this point we
do not neglect the oscillating terms in Eq.~(\ref{eq:H3rot}).

We conclude this section with a cautionary remark about the validity of the
spin Hamiltonian (\ref{eq:H_spin}) which describes only the lowest (acoustic) branch
of the magnon spectrum. Since YIG is a ferrimagnetic insulator with a rather large number of spins per unit cell, the magnon spectrum has also several high-energy (optical) branches \cite{Cherepanov93} which
are not taken into account  via the spin Hamiltonian (\ref{eq:H_spin}).
It turns out, however, that in thermal equilibrium at room temperature these optical magnons have a much lower occupancy than the
low-energy magnons, so that at  the energy scales probed in the experiment \cite{Noack19} we can safely neglect the optical magnons. In principle we cannot exclude the possibility that 
non-equilibrium scattering processes lead to a significant population of the optical magnons.
In fact, a recent calculation of the inverse spin-Hall voltage and the spin Seebeck effect in 
YIG by Barker and Bauer \cite{Barker16}
suggests that optical magnons can significantly contribute to spin transport.
On the other hand,  in Ref.~[\onlinecite{Barker16}] is is also shown  that the inclusion of the 
optical magnons does not qualitatively change the predicted inverse spin-Hall voltage.
Since in the present work we do not attempt to calculate the absolute size of the 
inverse spin-Hall voltage but  consider only the
magnon density (which is expected to
be proportional to the inverse spin-Hall voltage),  for our purpose it is sufficient
to work with the effective low-energy spin Hamiltonian (\ref{eq:H_spin}).
The high-energy magnon bands can at least partially be taken into account
by considering the parameters in Eq.~(\ref{eq:H_spin}) as effective  quantities which include 
renormalization  effects due to the
optical magnon bands. This argument is further strengthened by the fact that
the Hamiltonian (\ref{eq:H_spin}) correctly describes
the dynamics of non-equilibrium magnon condensation in YIG \cite{Rueckriegel15}.

\section{\label{sec:kinetic_eq}
Collisionless kinetic equations and S-theory with phenomenological damping}

Before deriving in Sec.~\ref{sec:col_int_der} kinetic equations for the 
distribution functions of magnons in YIG including the  
relevant collision integrals,
it is instructive to consider first the collisionless limit.
As recently pointed out in Ref.~[\onlinecite{Hahn20}],
for a complete description of the non-equilibrium time-evolution 
of the magnon distribution in YIG, we should take into account that in 
the presence of a time-dependent microwave field
the magnon annihilation operators can have a finite expectation value exhibiting a
non-trivial dynamics. In the rotating reference frame we define
 \begin{eqnarray}
\tilde{\psi}_\kv(t) &=& \langle \tilde{a}_\kv(t) \rangle =
    e^{i\omega_0t/2} \langle {a}_\kv(t) \rangle =
 e^{i\omega_0t/2}\psi_\kv (t ), 
 \hspace{7mm}
\end{eqnarray}
where the time-evolution is in the Heisenberg picture and $\langle\dots\rangle$ denotes to the non-equilibrium statistical average.
In addition, we should consider the time-evolution of the
connected diagonal- and off-diagonal distribution functions,
\begin{eqnarray}
n^c_\kv(t) &=& \langle \delta a^\dagger_\kv(t) \delta a_\kv(t)\rangle = \langle 
 \delta \tilde{a}^\dagger_\kv(t) \delta \tilde{a}_\kv(t)\rangle,\\
\tilde{p}^c _\kv(t) &=& \langle \delta \tilde{a}_{-\kv}(t) \delta \tilde{a}_\kv(t)\rangle = \mbox{e}^{i \omega_0 t} p_\kv(t),
\end{eqnarray}
where $\delta a_{\bd{k}} ( t ) = a_{\bd{k}} ( t ) - \langle a_{\bd{k}} ( t ) \rangle 
=   a_{\bd{k}} ( t ) - \psi_{\bd{k}} ( t )$.
 Note that  the phase factors $e^{\pm i\omega_0t/2}$ generated by the transformation to the rotating reference frame cancel
in  the diagonal distribution function $n^c_\kv(t)$.

\subsection{Collisionless kinetic equations}
\label{sec:collisionless}

The equations of motion for the distribution functions can be derived from the Heisenberg equations of motion for the operators in the rotating reference frame,
\begin{subequations}
\begin{eqnarray}
 i \partial_t \tilde{a}_{\bd{k}} &=& \bigl[ \tilde{a}_{\bd{k}} , \tilde{\cal{H}} ( t ) \bigr], \\
 i \partial_t \tilde{a}^{\dagger}_{\bd{k}} &=& \bigl[ \tilde{a}^{\dagger}_{\bd{k}} , \tilde{\cal{H}} ( t ) \bigr].
\end{eqnarray}
\end{subequations}
To begin with, let us approximate the magnon Hamiltonian 
by its quadratic part $\tilde{\mathcal{H}}_2$ neglecting all magnon-magnon interactions. In this approximation~\cite{Hahn20},
 \begin{subequations}
  \label{eq:kinfree}
\begin{eqnarray}
 \partial_t n_\kv^c + i \left[V_\kv \left(\pvf{\kv}\right)^* - V_\kv^* \pvf{\kv}\right] &=& 0, \label{eq:dgl_n0} \\
 \partial_t \pvf{\kv} + 2 i E_\kv \pvf{\kv} + i V_\kv \left[2 n_\kv^c + 1\right] &=& 0, \label{eq:dgl_p0} \\
 \partial_t \vfv{\kv} + i E_\kv \vfv{\kv}  + i V_\kv \vfe{-\kv} &=& 0. \label{eq:dgl_a0}
\end{eqnarray}
 \end{subequations}
Unfortunately, these equations do not provide a satisfactory 
description of the experimental results of Ref.[\onlinecite{Noack19}]. 
In particular, 
in the strong pumping regime  $\left|V_\kv\right|>\left|E_\kv\right|$ 
these equations predict an exponential growth
of the magnon distributions\cite{Suhl57,Schloemann60,Kloss10}, whereas experimentally one observes a saturation for  sufficiently long times.
To describe this  saturation we have to take magnon-magnon interactions into account. 
This can be done by employing a time-dependent self-consistent 
Hartree-Fock approximation, which in this context is 
called S-theory \cite{Zakharov70,Zakharov74,Lim88,Lvov94,Rezende09}.
The kinetic equations (\ref{eq:kinfree}) are then replaced
by non-linear integro-differential equations, which 
in the rotating reference frame take again the form~\cite{Hahn20}
\begin{subequations}
\label{eq:dgl_no_col_int}
\begin{eqnarray}
 \partial_t n_\kv^c + i \left[\tilde{V}_\kv \left(\pvf{\kv}\right)^* - \tilde{V}_\kv^* \pvf{\kv}\right] &=& 0, \label{eq:dgl_n}\\
 \partial_t \pvf{\kv} + 2 i \tilde{E}_\kv \pvf{\kv} + i \tilde{V}_\kv \left[2 n_\kv^c + 1\right] &=& 0, \label{eq:dgl_p}\\
 \partial_t \vfv{\kv} + i \tilde{E}_\kv \vfv{\kv}  + i \tilde{V}_\kv \vfe{-\kv} &=& 0, 
\label{eq:dgl_a}
\end{eqnarray}
\end{subequations}
where the renormalized magnon energy $\tilde{E}_\kv$ and the renormalized pumping energy $\tilde{V}_\kv$ depend on the distribution functions as follows,
\begin{subequations}
 \label{eq:renen}
\begin{eqnarray}
 \tilde{E}_\kv &=& E_\kv + \frac{1}{N} \sum\limits_\qv T_{\kv, \qv} 
\Bigl( n^c_\qv + 
 \bigl| \tilde{\psi}_{\qv} \bigr|^2 \Bigr),\label{eq:E_r_ex}\\
 \tilde{V}_\kv &=& V_\kv + \frac{1}{2 N} \sum\limits_\qv S_{\kv, \qv} \left(\tilde{p}_\qv^c + 
\tilde{\psi}_{-\qv}\tilde{\psi}_{\qv}\right).
\label{eq:V_r_ex}
\end{eqnarray}
\end{subequations}
Here $T_{\kv,\qv}$ and $S_{\kv,\qv}$ are defined via the following matrix elements of 
magnon-magnon interaction vertices in Eq.~(\ref{eq:H4rot}),
\begin{subequations}
\begin{eqnarray}
T_{\kv,\qv} &=& \Gamma^{\bar{a}\bar{a}aa}_{-\kv,-\qv;\qv,\kv},\\
S_{\kv,\qv} &=& \Gamma^{\bar{a}\bar{a}aa}_{-\kv,\kv;-\qv,\qv}.
\end{eqnarray}
\end{subequations}
Note that in Eq.~(\ref{eq:renen}) we have dropped oscillating terms 
arising from the vertices of ${\tilde{\cal{H}}}_4 ( t )$ in Eq.~(\ref{eq:H4rot})
involving time-dependent factors of 
$e^{\pm i\omega_0t}$ and $e^{\pm 2i\omega_0t}$,
which is consistent within the
rotating-wave approximation.

\subsection{\label{sec:stat}
Stationary non-equilibrium distribution with phenomenological damping}

\begin{figure}
 \centering
 \includegraphics[clip=true,trim=20pt 180pt 50pt 200pt,width=\linewidth]{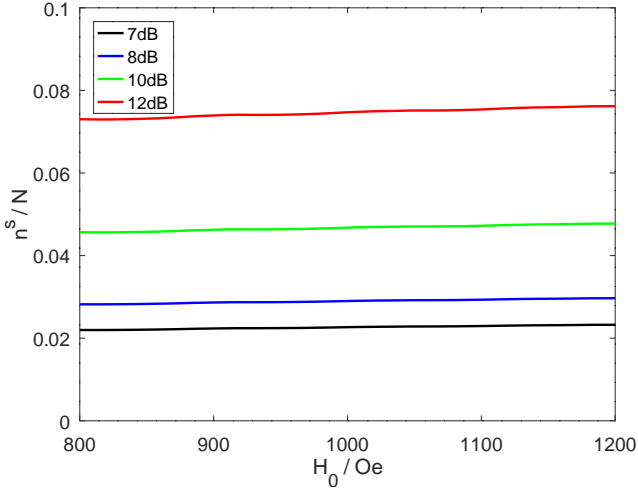}
 \caption{
Dependence of the magnon density $n^s / N = \sum_\kv n^s_\kv / N$ 
 on the external magnetic field strength $H_0$
in the stationary non-equilibrium state within S-theory given by Eq.~\eqref{eq:n_stat_S} for different pumping strengths. The maximum of $V_\kv$ was chosen to be larger than the relaxation rate $\gamma_\kv = 2.19 \times 10^{-3}$GHz.
To describe the experiment of Noack {\it{et al}} \cite{Noack19}
we have performed our calculations for 
a  thin YIG film  with  thickness $d=22.8 \, \mu$m 
(corresponding to  $N = 18422$)  subject to a  
microwave field with frequency $\omega_0=13.857$ GHz.
}
 \label{fig:mdS}
\end{figure}

In the experiment by Noack {\it{et al.}}\cite{Noack19} the magnetic-field dependence of the
magnon distribution in a stationary non-equilibrium state of a YIG sample subject to an oscillating microwave field is measured. Let us now try to
explain this experiment using a simple modification of the collisionless
kinetic equations (\ref{eq:dgl_no_col_int})
where we introduce (by hand) a phenomenological damping rate $\gamma_{\bd{k}}$. 
Note that without such a damping rate the solutions of the collisionless kinetic equations never reach a stationary non-equilibrium state \cite{Hahn20}.
In the rotating reference frame the equations of motion for the magnon operators including the phenomenological damping $\gamma_{\bd{k}}$ are
\begin{subequations}
 \begin{eqnarray}
  \partial_t\tilde{a}_\kv(t)&=& \left(- iE_\kv - \gamma_\kv\right)\tilde{a}_\kv-iV_\kv\tilde{a}_{-\kv}^\dagger, \\
  \partial_t\tilde{a}_\kv^\dagger(t)&=&\left( iE_\kv -\gamma_\kv\right) \tilde{a}_\kv^\dagger
 +iV_\kv^\ast \tilde{a}_{-\kv}.
 \end{eqnarray}
\end{subequations}
In Refs.~[\onlinecite{Zakharov70,Zakharov74}] it was argued that the damping selects the pair of magnon modes with momentum $\pm\kv$ that is characterized by the smallest damping to be the only significantly occupied modes, so that
the dynamics of these modes is
effectively decoupled from the other modes. Moreover, 
it is argued that, if initially other magnon modes are significantly occupied as well, after sufficiently long times only this single pair of magnon modes will survive.
This argument justifies the 
approximation of replacing the integrals defining the renormalized energies 
in Eq.~(\ref{eq:renen})
by a single  term where the loop momentum $\qv$ is 
equal the external momentum $\kv$,
\begin{subequations}
 \label{eq:EVS}
\begin{eqnarray}
 \tilde{E}_\kv &\approx& E_\kv + \frac{1}{N} T_{\kv, \kv}
 \Bigl( n^c_\kv + 
 \bigl| \tilde{\psi}_{\kv} \bigr|^2 \Bigr),\label{eq:E_r_dc}\\
 \tilde{V}_\kv   &\approx& V_\kv + \frac{1}{2 N} S_{\kv, \kv} \left(\tilde{p}_\kv^c + 
 \tilde{\psi}_{-\kv}\tilde{\psi}_{\kv}\right).
 \label{eq:V_r_dc}
\end{eqnarray}
\end{subequations}
Neglecting the expectation values of the magnon operators, Zakharov 
{\it{et al.}}~\cite{Zakharov70,Zakharov74} find
that the  stationary solution of the collisionless kinetic equations 
\eqref{eq:dgl_no_col_int} with additional damping is given by
\begin{subequations}
 \label{eq:stat_S}
\begin{eqnarray}
 n^s_\kv &=& N \frac{\sqrt{V_\kv^2 - \gamma_\kv^2}-\bigl|E_\kv\bigr|}{T_{\kv, \kv} + \frac{1}{2} S_{\kv, \kv}},
 \label{eq:n_stat_S}\\
 \tilde{p}^s_\kv &=& - n^s_\kv,
 \label{eq:p_stat_S}
\end{eqnarray}
\end{subequations}
provided the pumping is strong enough to compensate the losses due to damping,
\begin{equation}
 \left|V_\kv\right|>\left|\gamma_\kv\right|.
 \label{eq:ST_damp_cond}
\end{equation}
We shall refer to Eq.~(\ref{eq:stat_S}) as the stationary 
solution within S-theory.
Taking explicitly the expectation values of the magnon operators 
in Eq.~(\ref{eq:EVS})
into account yields the same result~\cite{Hahn20,footnoteamb},
\begin{subequations}
\begin{eqnarray}
 n^c_\kv+\left|\vfv{\kv}\right|^2 &=& n^s_\kv, \\
 \tilde{p}^c_\kv+ \vfv{\kv}^2 &=& - n^s_\kv.
\end{eqnarray}
\label{eq:n_stat_0}
\end{subequations}
In Fig.~\ref{fig:mdS} we plot the stationary magnon density 
$n^s = \sum_{\bd{k}} n_{\bd{k}}^s$ within S-theory obtained from Eq.~(\ref{eq:stat_S}) as a function of the external magnetic field assuming
a constant phenomenological relaxation rate $\gamma_{\bd{k}} = 2.08 \times 10^{-3}$GHz.
\begin{figure}
 \centering
 \includegraphics[clip=true,trim=20pt 180pt 50pt 200pt,width=\linewidth]{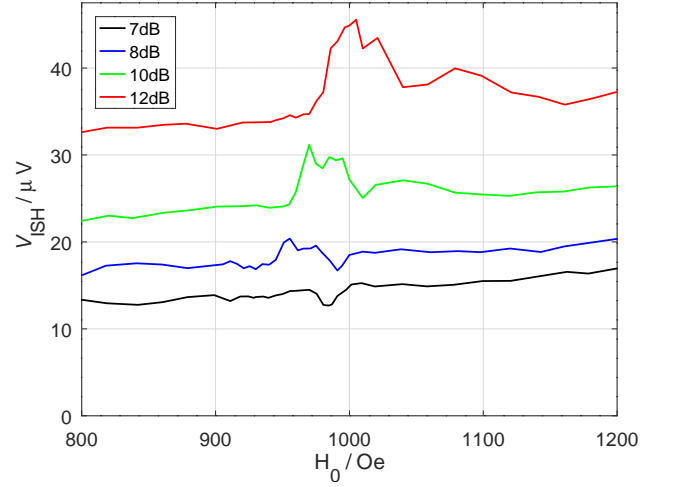}
 \caption{
Experimental results for the inverse spin-Hall voltage $V_{\rm{ISH}}$
reproduced from Fig.~4~a) of Ref.~[\onlinecite{Noack19}].
}
 \label{fig:experiment}
\end{figure}
For comparision, we reproduce in Fig.~\ref{fig:experiment} the experimental results for the
inverse spin-Hall effect voltage from Fig.~4 a) of Ref.~[\onlinecite{Noack19}], which is expected to be proportional to the density of pumped magnons.
Obviously, in a certain range of magnetic fields the experimental data exhibit
characteristic features which are missed by S-theory, 
which explains only 
the average linear growth of the observed magnon density with increasing 
magnetic field.
An obvious reason for the failure of S-theory
is that the phenomenological damping introduced by hand 
neither takes into account the kinematic constraints nor the 
microscopic magnon dynamics responsible for the dissipative effects which are essential 
for the emergence of a stationary non-equilibrium state in the pumped magnon gas.
For a satisfactory  explanation of the experimental data \cite{Noack19}
reproduced in the lower part of Fig.~\ref{fig:experiment} we should therefore
use kinetic equations with  microscopically derived collision integrals
describing the relevant scattering processes.
The collisionless kinetic equations (\ref{eq:kinfree}) are then replaced by
\begin{subequations}
\label{eq:dgl_ci}
\begin{eqnarray}
 \partial_t n_\kv^c(t) + i \left[\tilde{V}_\kv(t) \left(\pvf{\kv}(t)\right)^* - \tilde{V}_\kv^*(t) \pvf{\kv}(t)\right] &=& I^n_\kv(t), \nonumber\\
 \label{eq:dgl_n_ci}\\
 \partial_t \pvf{\kv}(t) + 2 i \tilde{E}_\kv(t) \pvf{\kv}(t) + i \tilde{V}_\kv(t) \left[2 n_\kv^c(t) + 1\right] &=& I^p_\kv(t), \nonumber\\
 \label{eq:dgl_p_ci} \\
 \partial_t \vfv{\kv}(t) + i \tilde{E}_\kv(t) \vfv{\kv}(t)  + i \tilde{V}_\kv(t) \vfe{-\kv}(t) &=& I^\psi_\kv(t), \nonumber\\
& &
\end{eqnarray}
\end{subequations}
where all interactions beyond S-theory are taken into account via three types of 
collision integrals $I^n_\kv(t)$, $I^p_\kv(t)$, and $I^\psi_\kv(t)$. These collision integrals should be derived from the Hamiltonian~\eqref{eq:H_1/S},
including the cubic  part $\mathcal{\tilde{H}}_3 (t)$ which determines the 
damping  to leading order in the small parameter $1/S$.
In spite of many decades of theoretical research on pumped magnon gases,\cite{Suhl57,Schloemann60,Zakharov70,Zakharov74,Vinikovetskii79,Cherepanov93,Araujo74,Tsukernik75,Lavrinenko81,Zvyagin82,Zvyagin85,Lim88,Kalafati89,Lvov94,Zvyagin07,Rezende09,Kloss10,Safonov13,Slobodianiuk17,Hahn20} a complete derivation of the relevant collision integrals
 $I^n_\kv(t)$, $I^p_\kv(t)$, and $I^\psi_\kv(t)$ and the subsequent numerical solution of the
resulting kinetic equations cannot be found in the literature. In the rest of this work we 
will solve this technically very complicated problem using an unconventional
approach to non-equilibrium many-body systems developed by
J. Fricke \cite{Fricke97} which we review in Appendix~B.

Before deriving in the following section explicit microscopic expressions for the
collision integrals in Eq.~(\ref{eq:dgl_ci}) let us generalize the construction of a stationary
solution with phenomenological damping discussed above by assuming that the
collision integrals are of the form
\begin{subequations}
\label{eq:gamma_approx}
\begin{eqnarray}
 I^n_\kv(t)&=&\gamma^n_\kv n_\kv(t), \\
 I^p_\kv(t)&=&\gamma^p_\kv \tilde{p}_\kv(t),
\end{eqnarray}
\end{subequations}
where $\gamma^n_\kv$ and $\gamma^p_\kv$ are assumed to be constant in time and independent
of  the magnon distribution functions.
For simplicity we assume that the expectation values of the magnon operators are negligible and
set $I^{\psi}_{\bd{k}} ( t ) =0$. 
In this case the stationary non-equilibrium solution of
Eq.~(\ref{eq:dgl_ci})
can easily be obtained  analytically. The imaginary part of $\gamma^p_\kv$ can be grouped together with the renormalized magnon energy $\tilde{E}_\kv$ and we therefore modify the expression for the renormalized magnon energy as follows,
\begin{equation}
 \tilde{E}_\kv=E_\kv-\frac{1}{2}\mathrm{Im} \gamma^p_\kv  + \frac{1}{N} \sum\limits_\qv T_{\kv,\qv} n_\qv(t).
\end{equation}
For $|V_\kv| > \frac{1}{4}\gamma^n_\kv\mathrm{Re} \gamma^p_\kv $
the stationary non-equilibrium solution of Eq.~(\ref{eq:dgl_ci})
is then given by
\begin{subequations}
\label{eq:n_stat}
\begin{eqnarray}
 n^s_\kv&=&\sqrt{\frac{\mathrm{Re} \gamma^p_\kv }{\gamma^n_\kv}}\left|\tilde{p}_\kv\right|,
 \label{eq:n_stat_cci}\\
 \tilde{p}^s_\kv&=&-\left(\sqrt{1-\frac{\gamma^n_\kv\mathrm{Re} \gamma^p_\kv }{4V_\kv^2}}+i\sqrt{\frac{\gamma^n_\kv\mathrm{Re} \gamma^p_\kv }{4V_\kv^2}}\right) \left|\tilde{p}_\kv\right|,\nonumber\\
 \label{eq:p_stat_cci}\\
 \left|\tilde{p}^s_\kv\right|&=&N\frac{\sqrt{V_\kv^2-\frac{1}{4}\gamma^n_\kv\mathrm{Re} \gamma^p_\kv }-|E_\kv|\sqrt{\gamma^n_\kv/\mathrm{Re} \gamma^p_\kv }}{T_{\kv,\kv}+\frac{1}{2}S_{\kv,\kv}}. \nonumber\\
\end{eqnarray}
\end{subequations}
Note that for $\frac{1}{2}\gamma^n_\kv = \frac{1}{2}\gamma^p_\kv \equiv \gamma_\kv$ 
we recover the stationary solution within conventional S-theory~\cite{Zakharov70,Zakharov74}
given in Eqs.~(\ref{eq:stat_S}).
Contrary to the case without collision integrals, the result for non-vanishing expectation values $\vfv{\kv}$ differs as the collision integrals cannot be written in the form $\gamma^n_\kv (n_\kv+|\vfv{\kv}|^2)$.

\section{\label{sec:col_int_der}
Collision Integrals}

In this section we present a microscopic derivation of the
collision integrals $I^n_\kv(t)$, $I^p_\kv(t)$, and $I^\psi_\kv(t)$
appearing in the kinetic equations~(\ref{eq:dgl_ci}).
Given the fact that for YIG the effective spin $S \approx 14$ is rather large \cite{Kreisel09},
we work to leading order in $1/S$ where only 
the cubic part $\mathcal{\tilde{H}}_3 ( t )$ of the Hamiltonian 
in Eq.~(\ref{eq:H3rot}) has to be taken into account. The assumption that
the experimentally observed fine structure of the inverse spin-Hall signal 
shown in Fig.~\ref{fig:experiment}
can be explained with the help of the scattering processes described by the cubic vertices
contained in  $\mathcal{\tilde{H}}_3 ( t )$ is also supported by the fact that
the peaks and dips of the observed signal as a function of the magnetic field agree with the
points where the 
 splitting processes (in which one magnon is absorbed and two magnons are emitted) 
and the confluence processes (in which two magnons are absorbed and one magnon is emitted)
described by the vertices in $\mathcal{\tilde{H}}_3 ( t )$
become kinematically possible \cite{Noack19}.
Note that a finite  cubic part $\mathcal{\tilde{H}}_3 (t)$ of the magnon Hamiltonian
arises 
entirely from dipole-dipole interactions. The corresponding scattering processes
conserve energy and momentum, but do not conserve the number of 
magnons \cite{Filho00}.
As we do not expect magnon-phonon interactions, magnon-defect interactions, and interactions with thermal optical magnons to be responsible for the effect observed in the experiment \cite{Noack19} we neglect these interactions.

In principle, the collision integrals can be derived using 
the Keldysh formalism \cite{Kamenev11}. However the Keldysh formalism has the disadvantage that it produces two-time correlations,
whereas in our case we are only interested in equal-time correlations.
Although the reduction of two-time correlations to equal-time correlations 
can be achieved by means of standard methods such as the 
generalized Kadanoff-Baym-Ansatz~\cite{Lipavsky86}, in view of the complexity of the
collision integrals for YIG  we find it more efficient to use a method
involving  only equal-time correlations at every step of the calculation.
We therefore use the method developed by J. Fricke \cite{Fricke97},
which allows us to to derive directly a hierarchy of coupled kinetic equations 
for  equal-time correlations and provides us with  
a systematic scheme for decoupling the correlations for arbitrary order.
To make this work self-contained, in Appendix~B we outline the main features of this method.

\subsection{Collision integrals due to cubic interaction vertices}

\begin{figure}[tb]
 \includegraphics[width=\linewidth]{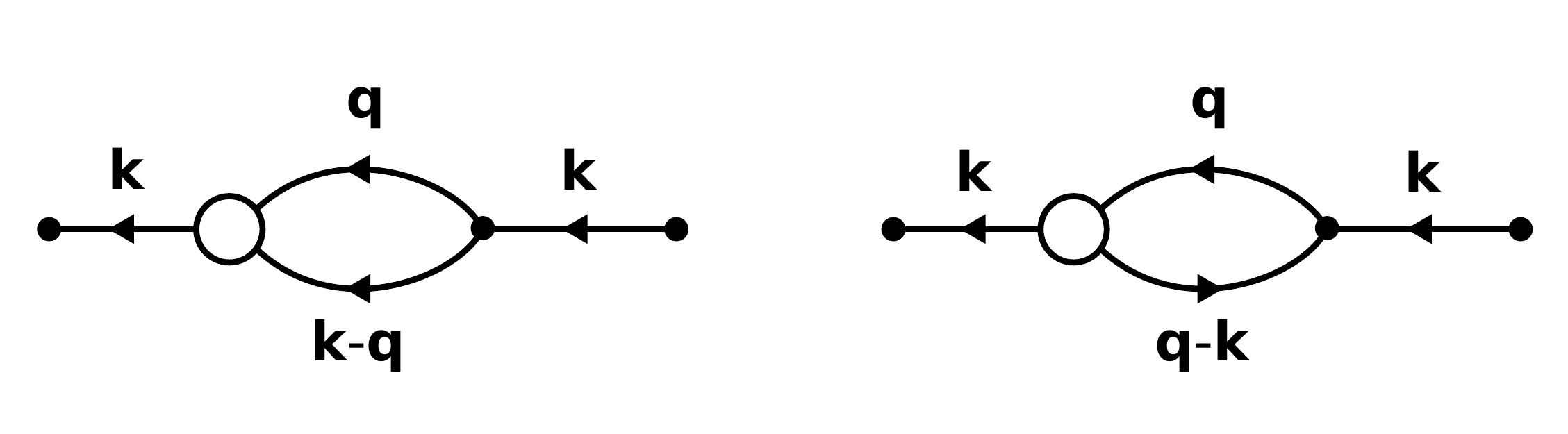}
 \caption{Diagrammatic representation of contributions to the 
collision integral $I^n_{\bd{k}} (t)$ given in \eqref{eq:n_3pc} which determine
the time-evolution of the connected diagonal distribution function $n^c_\kv (t )$. For simplicity 
we do not draw the two conjugated diagrams obtained by flipping the direction of each arrow corresponding to the complex conjugated terms in Eq.~(\ref{eq:n_3pc}).
The symbols have the following meaning:
Outgoing arrows represent creation operators, incoming arrows
represent  annihilation operators, and  the black dots represent external or interaction vertices.  
The left diagram contains two external vertices and the interaction vertex 
$\gam{\kv}{\qv}{\kqv}$; the empty circle (correlation bubble) represents the 
correlation $\langle \tilde{a}_\qv^\dagger \tilde{a}_\kqv^\dagger \tilde{a}_\kv\rangle^c$.
As the lines between the correlation bubble and the interaction vertex in the left diagram form a pair of equivalent lines 
we have to insert a  prefactor of $1 / 2$ in front of the first
vertex in Eq.~(\ref{eq:n_3pc}).
The right diagram contains the vertex $\Gamma^{\bar{a}\bar{a}a}_{\kv, \qkv; \qv}$ and the correlation $\langle \tilde{a}_\qv^\dagger \tilde{a}_\qkv \tilde{a}_\kv \rangle^c$.}
 \label{fig:diag_n1}
\end{figure}

\begin{figure}[tb]
 \centering
 \includegraphics[clip=true,trim=0pt 720pt 320pt 0pt,scale=0.4]{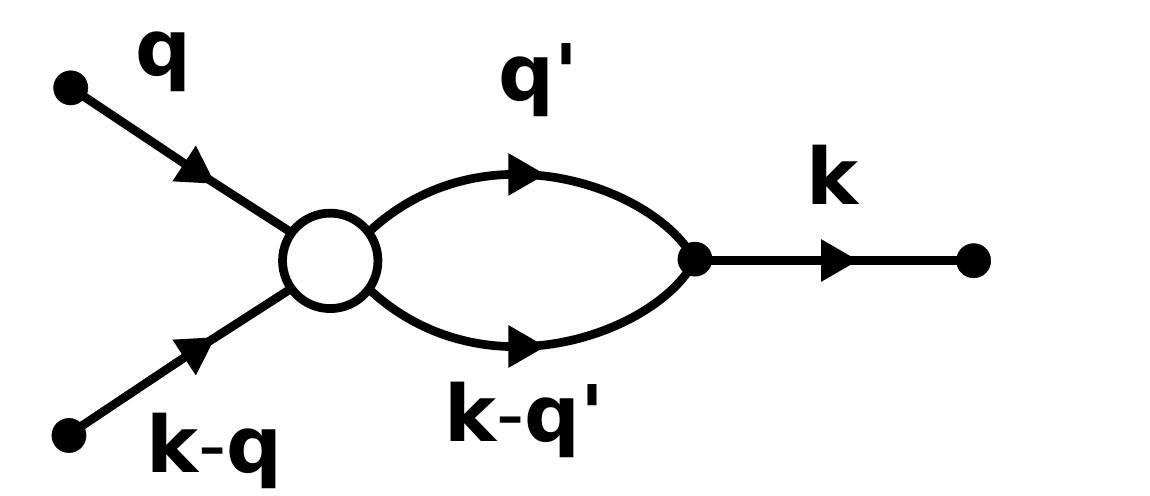}
 \caption{One of the diagrams contributing to the equation of motion of the three-point correlation $\langle \tilde{a}^\dagger_\qv \tilde{a}^\dagger_\kqv \tilde{a}_\kv \rangle^c$. 
This diagram, which  corresponds to the term explicitly written out in Eq.~\eqref{eq:3p1_dgl},
 contains the interaction vertex $\Gamma^{\bar{a}\bar{a}a}_{\qv', \kv-\qv'; \kv}$ and the four-point
correlation $\langle \tilde{a}^\dagger_\qv \tilde{a}^\dagger_\kqv \tilde{a}_{\qv'} \tilde{a}_{\kv-\qv'} \rangle^c$. As the lines between the correlation bubble and the interaction vertex are a pair of 
equivalent lines this diagram should be weighted by an extra factor of $1 / 2$.}
 \label{fig:diag_3p1}
\end{figure}

Consider first the diagonal collision integral $I^n_{\bd{k} } ( t )$
appearing in the kinetic equation (\ref{eq:dgl_n_ci}) for the connected part $n^c_{\bd{k}} ( t )$
of the diagonal magnon distribution.
Using the method developed in Ref.~[\onlinecite{Fricke97}] (which we review in Appendix~B) and omitting for simplicity the time-arguments, we find
\begin{eqnarray}
 I^n_\kv (t) =  \frac{i}{\sqrt{N}} &\sum\limits_\qv \left[\frac{1}{2} \gam{\kv}{\qv}{\kqv} e^{-i\omega_0t/2} \langle \tilde{a}_\qv^\dagger \tilde{a}_\kqv^\dagger \tilde{a}_\kv \rangle^c - \mathrm{c.c.} \right.\nonumber\\
 &\left. + (\gam{\qv}{\qkv}{\kv})^* e^{i\omega_0t/2} \langle \tilde{a}_\qv^\dagger \tilde{a}_\qkv \tilde{a}_\kv \rangle^c - \mathrm{c.c.}\right], \nonumber\\
 \label{eq:n_3pc}
\end{eqnarray}
where we have used momentum conservation to carry out one of the summations.
Here  $\langle \tilde{a}_\qv^\dagger \tilde{a}_\kqv^\dagger \tilde{a}_\kv \rangle^c$ and
$ \langle \tilde{a}_\qv^\dagger \tilde{a}_\qkv \tilde{a}_\kv \rangle^c$ are connected equal-time correlations involving
three magnon operators.  In the graphical representation of Eq.~(\ref{eq:n_3pc})
shown Fig.~\ref{fig:diag_n1} 
these correlations are represented by empty circles with three external legs 
(correlation bubbles).
Note that the diagrams shown in Fig.~\ref{fig:diag_n1}
differ from Feynman diagrams as they represent contributions to the differential equations for the correlations at a fixed time. 
Next, we express the three-point correlations in Eq.~(\ref{eq:n_3pc})
in terms of the four-point correlations using the equation of motion.
As a representative example, let us consider  the correlation
 $\langle \tilde{a}^\dagger_\qv \tilde{a}^\dagger_\kqv \tilde{a}_\kv \rangle^c$ in the first term
on the right-hand side of Eq.~(\ref{eq:n_3pc}) and explicitly evaluate
only the diagram shown in Fig.~\ref{fig:diag_3p1}. The other terms entering the equation of motion corresponding to the remaining diagrams have the same form and are represented by the dots
in Eqs.~(\ref{eq:3p1_dgl})--\ref{eq:obtain_cI}) below.
 The calculations leading to the collision integrals are analogous for all terms.
The equation of motion implies
\begin{widetext}
\begin{equation}
 \left[\frac{d}{dt} + i \left(\epsilon_\kv - \epsilon_\qv - \epsilon_\kqv\right)\right] \langle \tilde{a}^\dagger_\qv \tilde{a}^\dagger_\kqv \tilde{a}_\kv \rangle^c = -  \frac{i}{\sqrt{N}} \sum\limits_{\qv'} \left[
 \frac{1}{2} \left(\gam{\kv}{\qv'}{\kv-\qv'}\right)^* e^{i\omega_0t/2} \langle \tilde{a}^\dagger_\qv \tilde{a}^\dagger_\kqv \tilde{a}_{\qv'} \tilde{a}_{\kv-\qv'} \rangle^c + ... \right] .
 \label{eq:3p1_dgl}
\end{equation}
Integrating Eq.~(\ref{eq:3p1_dgl}) over the time we obtain
\begin{equation}
 \langle \tilde{a}^\dagger_\qv \tilde{a}^\dagger_\kqv \tilde{a}_\kv \rangle^c = - \frac{i}{\sqrt{N}} \sum\limits_{\qv'} \left[\frac{1}{2} \int\limits_{t_0}^t dt' \mathrm{e}^{-i \left(\epsilon_\kv - \epsilon_\qv - \epsilon_\kqv\right)\left(t - t'\right)} \left(\gam{\kv}{\qv'}{\kv-\qv'}\right)^* e^{i\omega_0t/2} \langle \tilde{a}^\dagger_\qv \tilde{a}^\dagger_\kqv \tilde{a}_{\qv'} \tilde{a}_{\kv-\qv'} \rangle^c + ... \right] .
 \label{eq:3p1}
\end{equation}
Finally, substituting Eq.~\eqref{eq:3p1} into Eq.~\eqref{eq:n_3pc} we obtain
\begin{eqnarray}
 I^n_\kv (t) &=&  \frac{1}{N} \sum\limits_{\qv,\qv'} \left[\frac{1}{2} \int\limits_{t_0}^t dt' \cos\left[\left(\epsilon_\kv - \epsilon_\qv - \epsilon_\kqv\right)\left(t - t'\right)\right] \gam{\kv}{\qv}{\kqv}\left(\gam{\kv}{\qv'}{\kv-\qv'}\right)^* \langle \tilde{a}^\dagger_\qv \tilde{a}^\dagger_\kqv \tilde{a}_{\qv'} \tilde{a}_{\kv-\qv'} \rangle^c  + ... \right] \nonumber\\
 \underrightarrow{t_0 \rightarrow - \infty}&& \frac{2 \pi}{N} \sum\limits_{\qv,\qv'} \left[\frac{1}{2} \delta\left(\epsilon_\kv - \epsilon_\qv - \epsilon_\kqv\right) \gam{\kv}{\qv}{\kqv}\left(\gam{\kv}{\qv'}{\kv-\qv'}\right)^* \langle \tilde{a}^\dagger_\qv \tilde{a}^\dagger_\kqv \tilde{a}_{\qv'} \tilde{a}_{\kv-\qv'} \rangle^c  + ... \right],
 \label{eq:obtain_cI}
\end{eqnarray}
\end{widetext}
where in the last step we have taken the limit $t_0\rightarrow-\infty$ and the dots denote the contributions of the other diagrams.
The other terms entering this equation represented 
by the dots are of the same form. Note that the terms with two annihilation operators
or two creation operators within the two-particle correlations are complex.
Therefore, there appears an exponential function with imaginary valued argument instead of the cosine function leading in the thermodynamic limit to a term of the same form as in Eq.~\eqref{eq:obtain_cI} without the factor of two. 
In this way all terms entering the equation of motion for the one-particle distribution functions can be obtained from the diagrams. A complete list of all diagrams contributing to
the equation of motion of the three-point correlations
$\langle \tilde{a}^\dagger_\qv \tilde{a}^\dagger_\kqv \tilde{a}_\kv \rangle^c $ 
and $\langle \tilde{a}_\qv^\dagger \tilde{a}_\qkv \tilde{a}_\kv \rangle^c$
is shown in Fig.~\ref{fig:diag_2a_1c} of Appendix~C.

\begin{figure}[tb]
 \centering
 \includegraphics[width=\linewidth]{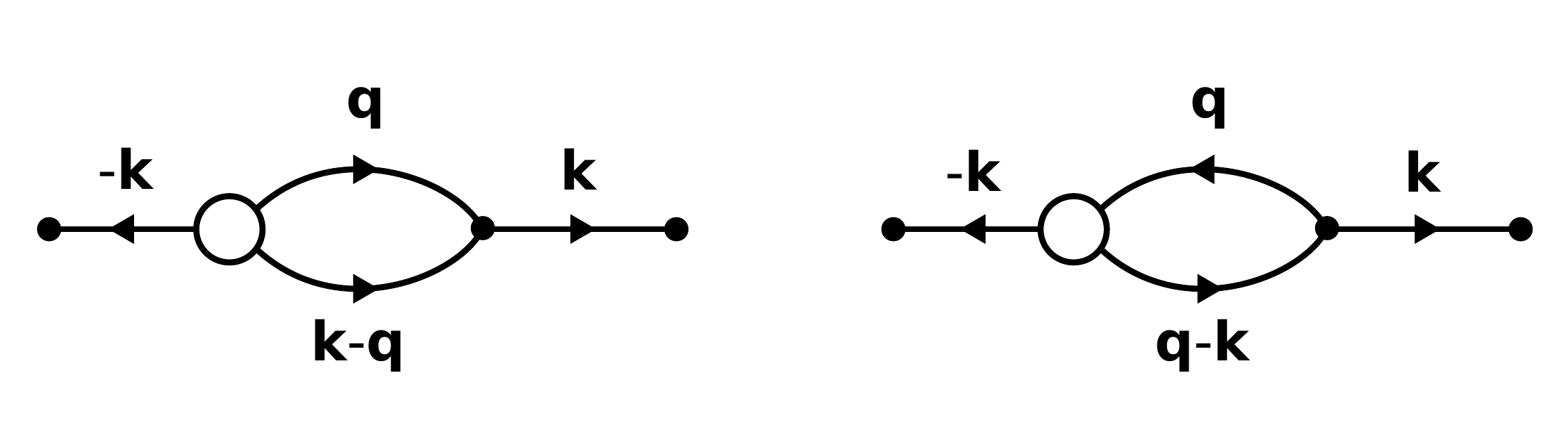}
 \caption{The two diagrams contributing to the time-evolution of the off-diagonal distribution function $p_\kv = \langle \tilde{a}_{-\kv} \tilde{a}_\kv \rangle$. The diagrams correspond to the two terms 
on the right-hand side of Eq.~\eqref{eq:p_3pc}. The left diagram contains the interaction vertex $\Gamma^{\bar{a}\bar{a}a}_{\qv, \kqv; \kv}$ and the correlation $\langle \tilde{a}_{-\qv} \tilde{a}_\qkv \tilde{a}_{-\kv} \rangle^c$. The left diagram should be multiplied by a factor of $1/2$ because
the lines between the correlation bubble and the vertex are equivalent.
 The right diagram contains the vertex $\gam{\qkv}{\qv}{\kv}$ and the correlation $\langle \tilde{a}^\dagger_\qv \tilde{a}_\qkv \tilde{a}_{-\kv} \rangle^c$.}
 \label{fig:offdiag_p1}
\end{figure}

The approach outlined above can also be used to obtain the off-diagonal collision integral
$I^p_{\bd{k}} ( t )$ in the kinetic equation (\ref{eq:dgl_p_ci}) 
for the off-diagonal distribution function $\pvf{\kv} ( t )$.
 In this case there are only two diagrams containing the relevant vertices shown
in Fig.~\ref{fig:offdiag_p1}. The corresponding expression for the
off-diagonal collision integral is
\begin{eqnarray}
 I^p_\kv (t) = -i \frac{1}{\sqrt{N}} \sum\limits_\qv &\left[\frac{1}{2} \left(\gam{\kv}{\qv}{\kqv}\right)^* e^{i\omega_0t/2} \langle \tilde{a}_{-\qv} \tilde{a}_\qkv \tilde{a}_{-\kv} \rangle^c \right.\nonumber\\
 &\left. + \gam{\qkv}{\qv}{\kv} e^{-i\omega_0t/2} \langle \tilde{a}^\dagger_\qv \tilde{a}_\qkv \tilde{a}_{-\kv} \rangle^c \right]. \nonumber\\
 \label{eq:p_3pc}
\end{eqnarray}
The correlation $\langle \tilde{a}_{-\qv} \tilde{a}_\qkv \tilde{a}_{-\kv} \rangle^c$ in the first term leads 
for large times to a delta function of the form $\delta\left(\epsilon_\kv + \epsilon_\qv + \epsilon_\kqv\right)$.
Keeping in mind that the magnon dispersion
 $\epsilon_\kv $ is positive for all momenta,  
this term does not contribute to the off-diagonal collision integral $I^p_\kv (t )$ for large times. 
The diagrams contributing to the correlation
$\langle \tilde{a}^\dagger_\qv \tilde{a}_\qkv \tilde{a}_{-\kv} \rangle^c$ have already been discussed in the context of
the diagonal collision integral 
$I^n_{\bd{k}} ( t )$, see  Fig.~\ref{fig:diag_2a_1c} in Appendix~C.
Finally, the collision integral $I^{\psi}_\kv (t )$ 
entering the kinetic equation (\ref{eq:dgl_ci}) for the expectation values 
 $\tilde\psi_\kv$
of the magnon operators vanishes,
\begin{equation}
 I^\psi_\kv (t )  = 0,
\end{equation}
because there is no diagram contributing to the time-evolution of $\tilde\psi_\kv (t)$ that is quadratic in the three-point vertices.

\subsection{Decoupling of the equations of motion for the
connected correlations}

So far, we have expressed the contributions to the collision integrals
involving the various types of three-point vertices
in terms of connected four-point correlations.
The next step is to decouple the hierarchy of equations of motion by replacing the connected four-point correlations by one-point and connected two-point correlations. Keeping in mind
that the only non-vanishing distribution functions are $n^c_\kv$, $\pvf{\kv}$, and $\vfv{\kv}$ we find
\begin{eqnarray}
 \langle \tilde{a}^\dagger_\kv \tilde{a}_\kv \tilde{a}^\dagger_\qv \tilde{a}_\qv \rangle^c &=& -1! \langle \tilde{a}^\dagger_\kv \tilde{a}_\kv \rangle \langle \tilde{a}^\dagger_\qv \tilde{a}_\qv \rangle + 2! \langle \tilde{a}^\dagger_\kv \tilde{a}_\kv \rangle \langle \tilde{a}^\dagger_\qv \rangle \langle \tilde{a}_\qv \rangle \nonumber\\
 && + 2! \langle \tilde{a}^\dagger_\kv \rangle \langle \tilde{a}_\kv \rangle \langle \tilde{a}^\dagger_\qv \tilde{a}_\qv \rangle - 3! \langle \tilde{a}^\dagger_\kv \rangle \langle \tilde{a}_\kv \rangle \langle \tilde{a}^\dagger_\qv \rangle \langle \tilde{a}_\qv \rangle \nonumber\\
 &=& -n^c_\kv n^c_\qv + n^c_\kv \left|\psi_\qv\right|^2 + n^c_\qv \left|\psi_\kv\right|^2 \nonumber\\
 &&- 3 \left|\psi_\kv\right|^2 \left|\psi_\qv\right|^2 .
 \label{eq:expansion}
\end{eqnarray}
Analogous expressions can be written down  for the other four-point correlations, so that
the collision integrals can be expressed in terms of the
two types of two-point correlations $n^c_{\bd{k}} ( t)$ and
$\tilde{p}^c_{\bd{k}} ( t )$ and the non-equilibrium expectation values $\psi_{\bd{k}} ( t )$
of the magnon operators. It is convenient to decompose the collision integrals as
\begin{subequations}
 \begin{eqnarray}
  I^p_\kv  = I^p_{\kv,\mathrm{in}}-I^p_{\kv,\mathrm{out}},\\
  I^p_\kv = I^p_{\kv,\mathrm{in}}-I^p_{\kv,\mathrm{out}},
 \end{eqnarray}
\end{subequations}
where $I_{\kv, \mathrm{in}}$ is the in-scattering or arrival term,  
and $I_{\kv, \mathrm{out}}$ is the out-scattering or departure term. 
The explicit expressions for the various contributions to the
collision integrals are rather cumbersome and are given in
Eqs.~\eqref{eq:In_in} - \eqref{eq:Ip_out} of Appendix~C. 
Within the rotating-wave approximation the fast oscillating terms containing  factors of
 $\mathrm{e}^{\pm i \omega_0 t}$ should be neglected to be consistent with a similar approximation
in  the renormalized magnon dispersion $\tilde{E}_\kv$ and the pumping energy $\tilde{V}_\kv$.

\section{\label{sec:md_cv}
Explanation of the magnetic field dependence of the inverse spin-Hall signal in YIG}

Having derived explicit expressions for the collision integrals
$I^n_{\bd{k}} ( t )$ and $I^p_{\bd{k}} ( t )$ we can now 
construct stationary solutions of the kinetic equations (\ref{eq:dgl_ci})
and determine the non-equilibrium magnon distribution which is proportional to the
inverse spin-Hall signal observed  in the experiment.\cite{Noack19}
As discussed in Sec.~\ref{sec:stat}, in order to understand the magnetic 
field dependence of the inverse spin-Hall signal we need a  microscopic 
understanding of the momentum-dependent magnon damping. In this section we first calculate the magnon damping
in thermal equilibrium which we need in the subsequent calculation of the
collision integrals.
We then present an approximate solution of the
kinetic equations (\ref{eq:dgl_ci})
with microscopic collision integrals derived in Sec.~\ref{sec:col_int_der}
and obtain excellent agreement with the experiment \cite{Noack19}.

\subsection{Magnon damping in thermal equilibrium}
\label{sec:dampeq}

\begin{figure}[tb]
 \centering
 \begin{minipage}{0.49\linewidth}
  (a) \hspace{64pt}
 \end{minipage}
 \begin{minipage}{0.49\linewidth}
  (b) \hspace{64pt}
 \end{minipage}
 \includegraphics[width=0.9\linewidth,clip=true,trim=40pt 40pt 40pt 50pt]{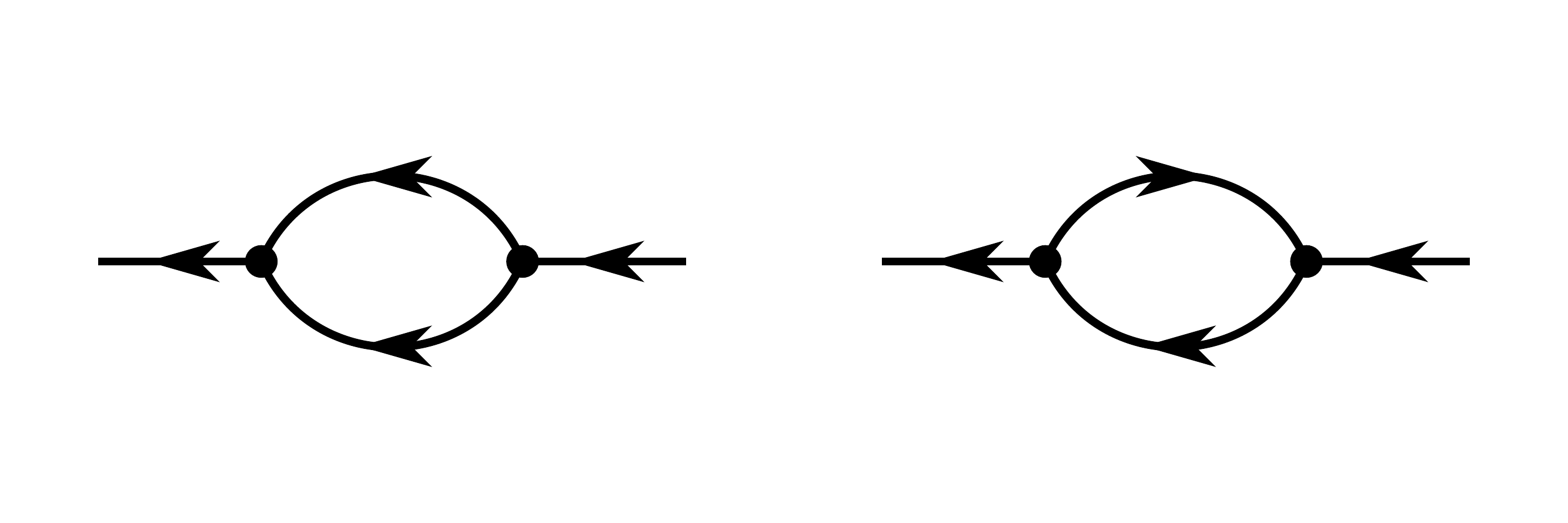}
 \caption{Feynman diagrams representing the contributions 
to the magnon self-energy which generate (a) the confluent and (b) the splitting
contributions to the 
magnon damping given in Eqs.~(\ref{eq:magconf}) and (\ref{eq:magsplit}).
Here the arrows represent the magnon propagators and the dots represent the cubic interaction vertices.}
\label{fig:Feynmandamp}
\end{figure}

\begin{figure}[tb]
  \centering
   \includegraphics[clip=true,trim=10pt 180pt 40pt 200pt,width=\linewidth]{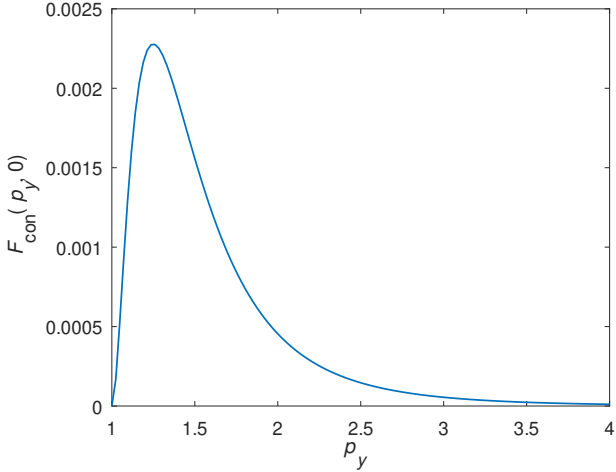}
  \caption{
Numerical evaluation of the function $F_{\rm con} ( p_y, 0 )$
defined in Eq.~(\ref{eq:Fcondef}) as a function of $p_y = k_y / \kappa$.
For large $p_y$ we find that $F_{\rm con} ( p_y , 0 ) \propto
1/ p_y^4$.
}
\label{fig:Fcon}
\end{figure}

In thermal equilibrium with temperature $T$ the normal magnon distribution is given by the
Bose-Einstein distribution
\begin{equation}
 n_\kv = \frac{1}{\mbox{e}^{\epsilon_\kv / T} - 1} .
 \label{eq:BoseEinstein}
\end{equation}
The magnon damping in equilibrium can then be 
obtained from the imaginary part of the magnon self-energy obtained within the 
imaginary-time (Matsubara) formalism. Alternatively, the magnon damping 
$\gamma^n_{\bd{k}}$
in equilibrium
can be obtained by writing the departure term of the collision integral as
\begin{equation}
 I^n_{\kv,\mathrm{out}} = \gamma^n_\kv n_\kv,
 \label{eq:Ingamma}
\end{equation}
where for simplicity we consider only the normal (diagonal) part $I^n_{\kv, \mathrm{out}}$
of the collision integral.
To simplify the explicit evaluation of the damping $\gamma^n_\kv$
let us assume that the momentum $\bd{k}$ is sufficiently large so that we can
neglect the effect of dipole-dipole interactions on the magnon dispersion. 
In this regime the long-wavelength magnon dispersion is determined by the
exchange interaction,
\begin{equation}
 \epsilon_\kv = \sqrt{A_\kv^2 - \left|B_\kv\right|^2} \approx \left|A_\kv\right| = h_0 + \rho \kv^2 ,
 \label{eq:eps2}
\end{equation}
with exchange stiffness 
 \begin{equation}
 \rho = J Sa^2.
 \end{equation}
 In the expressions for the magnon dispersion 
given in Appendix~A [see Eqs.~(\ref{eq:magdisp}) and (\ref{eq:magV})]
we can then set $B_\kv = 0$ and $V_\kv = 0$. 
According to  Ref.~[\onlinecite{Kreisel09}], for
the effective exchange energy in YIG is $J \approx 1.29 $K, the effective spin is 
$S \approx 14.2$, and the lattice constant is
$a \approx 12.376$ \AA.
The Bogoliubov transformation from Holstein-Primakoff bosons $b_{\bd{k}}$ to 
magnon operators $a_{\bd{k}}$ is then not necessary
so that we may identify the corresponding vertices,
 $\gam{\kv_1}{\kv_2}{\kv_3} = \Gamma_{\kv_1; \kv_2, \kv_3}^{\bar{b}bb}$.
Moreover, in the regime where the magnon dispersion is dominated by the
exchange energy we may neglect
the diagonal elements of the dipolar tensor $D_{\bd{k}}^{\alpha \beta}$
defined in Eq.~(\ref{eq:dipdef}). In the geometry
shown in Fig.~\ref{fig:setup} the only non-zero elements 
of the dipolar tensor are then
 $D^{yz}_\kv = D^{zy}_\kv$, see Eq.~(\ref{eq:Dyz}).
This greatly simplifies all quantities appearing in the kinetic equations for 
the magnon distribution. To get a rough estimate for the order of magnitude of the damping, let us also neglect the contributions from the off-diagonal distribution
$p_{\bd{k}} ( t )$ and the expectation values $\psi_{\bd{k}}$ 
of the magnon operators to the 
 collision integral $I^n_{\kv,\mathrm{out}}$ in Eq.~(\ref{eq:Ingamma}).
In this approximation we obtain 
\begin{equation}
 \gamma^n_\kv = \gamma^n_{\kv, \mathrm{con}} + \gamma^n_{\kv, \mathrm{split}},
\end{equation}
where the contribution from the confluent process is
\begin{eqnarray}
 \gamma^n_{{\bd{k}} , {\rm con}  }  & = & \frac{  \pi }{ N} \sum_{ \bd{q} }
\delta (  {\epsilon}_{\bd{k}}  - {\epsilon}_{\bd{k} - \bd{q} }  - {\epsilon}_{ \bd{q} }   )  
 \nonumber
 \\
& & \times  | \Gamma^{\bar{a} aa }_{  \bd{k} ; \bd{q} ,
 \bd{k} - \bd{q} } |^2 [ n_{\bd{q} }  + n_{\bd{k} - \bd{q}}  +1 ]  ,
 \label{eq:magconf}
 \end{eqnarray}
and the contribution from the splitting process is
 \begin{eqnarray}
\gamma^n_{{\bd{k}} , {\rm split}  }
 & = & 
 \frac{ 2 \pi}{ N} \sum_{ \bd{q} }
\delta (   {\epsilon}_{ \bd{k} } +  {\epsilon}_{\bd{q} - \bd{k} }     -   {\epsilon}_{\bd{q}}   )  
 \nonumber
 \\
& & \times  | \Gamma^{\bar{a} aa }_{  \bd{q} ; \bd{k} ,
 \bd{q} - \bd{k} } |^2  [ n_{\bd{q} - \bd{k}}  -   n_{\bd{q} }      ].
 \label{eq:magsplit}
 \end{eqnarray}
Note that these expressions can also be obtained directly from the diagonal part of the 
imaginary frequency magnon self-energy $\Sigma ( \bd{k} , i \omega )$ via analytic continuation,
 \begin{equation} 
 \gamma^n_{\bd{k}} = -  {\rm Im} \Sigma( \bd{k} , \epsilon_{\bd{k}} + i 0^+ ).
 \end{equation}
The Feynman diagrams for  the self-energy corrections associated with the 
confluence and the splitting processes are shown in Fig.~\ref{fig:Feynmandamp}.
For vanishing wavevector $\bd{k} =0$ the  confluent contribution has been carefully evaluated  
by  Chernyshev \cite{Chernyshev12}.
Here we are only interested in the range of wavevectors $\bd{k}$ where the magnon dispersion 
is dominated by the exchange energy so that it can be approximated by
$\epsilon_{\bd{k}} = h_0 + \rho \bd{k}^2$.
Keeping in mind that in our geometry the only non-vanishing matrix elements of the dipolar tensor are $D^{yz}_\kv = D^{zy}_\kv$ and using Eq.~(\ref{eq:Dyz}) we find
that the relevant cubic interaction vertex in Eqs.~(\ref{eq:magconf}) and (\ref{eq:magsplit}) is given by
\begin{eqnarray}
 \gam{\kv_1}{\kv_2}{\kv_3} &=& \Gamma^{\bar{b}bb}_{\kv_1, \kv_2, \kv_3} = \sqrt{\frac{S}{2}} \left(D^{zy}_{\kv_2} + D^{yz}_{\kv_3}\right) \nonumber\\
 &\approx& - \frac{\Delta}{\sqrt{2 S}} \left(\frac{k_{2 y} k_{2 z}}{k_2^2} + \frac{k_{3 y} k_{3 z}}{k_3^2}\right) ,
 \label{eq:gamma}
\end{eqnarray}
where the energy scale $\Delta$ associated with the dipolar interaction is defined
in Eq.~(\ref{eq:deltadipdef}).
Since the experiment\cite{Noack19} has been performed at room temperature which is large compared with the typical magnon energies, we may approximate the equilibrium magnon distribution
in Eqs.~(\ref{eq:magconf}) and (\ref{eq:magsplit}) by
a Rayleigh-Jeans distribution,
\begin{eqnarray}
 n_\qv &\approx& T / \epsilon_\qv, \; \; \; \; \; \;
 n_\kqv \approx T / \epsilon_\kqv .
\label{eq:n_approx}
\end{eqnarray}
Shifting the integration variable $\bd{q} = \bd{q}^{\prime} + \bd{k} /2$ in Eq.~(\ref{eq:magconf}), we obtain for the ratio
of the confluent magnon damping to the magnon energy at high temperatures,
 \begin{equation}
 \frac{ \gamma^{n}_{\bd{k} , {\rm con}}}{\epsilon_{\bd{k}}} = \frac{ T}{8 J} \left( \frac{ \Delta}{ h_0 S } \right)^2 \Theta ( 
 | \bd{k} |  - \kappa ) F_{\rm con} ( {\bd{k}} /  {\kappa } ),
 \label{eq:dampconf1}
 \end{equation}
where the threshold momentum $\kappa$ is defined by
 \begin{equation}
 \kappa^2 =  2 h_0 / \rho,
 \label{eq:kappa}
 \end{equation}
and the dimensionless function $F_{\rm con} ( \bd{p} )$ is defined via the following integral
 \begin{widetext}
 \begin{eqnarray}
 F_{\rm con} ( \bd{p} ) & = & \int_0^{ 2 \pi} \frac{ d \varphi}{ 2 \pi } 
 \frac{1}{ \Bigl[ 1 + \frac{1}{2} \left(  \bd{p} + \hat{\bd{q}}_{\varphi} \sqrt{ p^2 -1 } \right)^2 \Bigr]    
 \Bigl[ 1 + \frac{1}{2} \left(  \bd{p} - \hat{\bd{q}}_{\varphi} \sqrt{ p^2 -1 } \right)^2 \Bigr]
 }
 \nonumber
 \\
 & \times &
 \left[
 \frac{ \left( p_y + \sqrt{ p^2 -1 }  \cos \varphi    \right) \left(  p_z + \sqrt{ p^2 -1 }  \sin \varphi    \right)}{
 \left( \bd{p} + \hat{\bd{q}}_{\varphi} \sqrt{p^2 -1 } \right)^2}
 + 
 \frac{ \left( p_y -  \sqrt{ p^2 -1 }  \cos \varphi    \right) \left(  p_z - \sqrt{ p^2 -1 }  \sin \varphi    \right)}{
 \left( \bd{p} - \hat{\bd{q}}_{\varphi} \sqrt{p^2 -1 } \right)^2}
 \right]^2,
 \hspace{7mm}
 \label{eq:Fcondef}
 \end{eqnarray}
 \end{widetext}
where $\hat{\bd{q}}_{\varphi}= \bd{e}_y \cos \varphi + \bd{e}_z \sin \varphi$. At the threshold momentum
$\bd{k} = \kappa \hat{\bd{k}} $
this reduces to
 \begin{equation}
 F_{\rm con} ( \hat{\bd{k} } ) = \frac{16}{9} \hat{k}_y^2 \hat{k_z^2}.
\end{equation}
A numerical evaluation of $F_{\rm con} ( p_y, p_z=0 )$ is shown in Fig.~\ref{fig:Fcon}.
A rough estimate for the
order of magnitude of the confluent magnon damping for YIG at room temperature is
given by the prefactor in Eq.~(\ref{eq:dampconf1}), which yields\cite{Noack19}
 \begin{eqnarray}
 \frac{ \gamma^n_{\bd{k} , {\rm con}}}{\epsilon_{\bd{k}}} & \approx &
\frac{ T}{16 J} \left( \frac{ \Delta}{ h_0 S } \right)^2
 \nonumber
 \\
 & = & 
 \frac{ 290 K}{16 \times 1.29 K } \left( \frac{ 1750 \; G }{ 1000 \; G \times 14} \right)^2 
 \nonumber
 \\
 & = &  14 \times (0.125 )^2 \approx  0.22.
 \end{eqnarray}
This indicates that at room temperature the damping due to magnon confluence 
can be substantial.

Next, consider the contribution from the splitting process 
to the magnon damping in equilibrium represented by the diagram (b) in Fig.~\ref{fig:Feynmandamp}.
With the same approximations as above we obtain
 \begin{eqnarray}
  \frac{ \gamma^n_{\bd{k} , {\rm split}}}{\epsilon_{\bd{k}}} & = & 2 \pi T a^2 
 \int \frac{d^2 q}{(2 \pi)^2} 
 \frac{ \delta ( h_0 - 2 \rho \bd{k} \cdot \bd{q} ) }{ \epsilon_{\bd{q}} \epsilon_{ \bd{q} + \bd{k}} }
  \nonumber
 \\
 & \times & \frac{ \Delta^2}{2 S} 
 \left[ \hat{k}_y \hat{k}_z + \hat{q}_y \hat{q}_z \right]^2.
 \label{eq:dampsplit1}
 \end{eqnarray} 
Setting for simplicity $k_z = 0$, we see that 
the $\delta$-function enforces $q_y = q_y^0 = h_0 /( 2 \rho k_y )$.
The condition $| q_y^0 | \leq \pi / a $ then reduces to $ | k_y | > h_0 a /( 2 \pi \rho ) = \kappa^2 a /(4 \pi )$.
With $\kappa a  \ll 1$, it is clear that the splitting contribution to the magnon damping has a much lower threshold than the confluent contribution.
Using the quadratic approximation (\ref{eq:eps2})
for the magnon dispersion and the definition~\eqref{eq:kappa} of $\kappa$ 
we find that for parametrically pumped magnons with $\epsilon_\kv=\omega_0/2$
the condition $| \kv | > \kappa$ is satisfied for
\begin{equation}
h_0 < \frac{\omega_0}{6} ,
\label{eq:conf_quad}
\end{equation}
where the upper bound $\omega_0 / 6$ 
coincides with the magnetic field strength below which the confluent
damping process is kinematically possible.
On the other hand, for  $h_0 > \omega_0 /6$ the damping is dominated 
by the splitting processes.

\begin{figure}[tb]
 \centering
 \includegraphics[clip=true,trim=10pt 180pt 40pt 200pt,width=\linewidth]{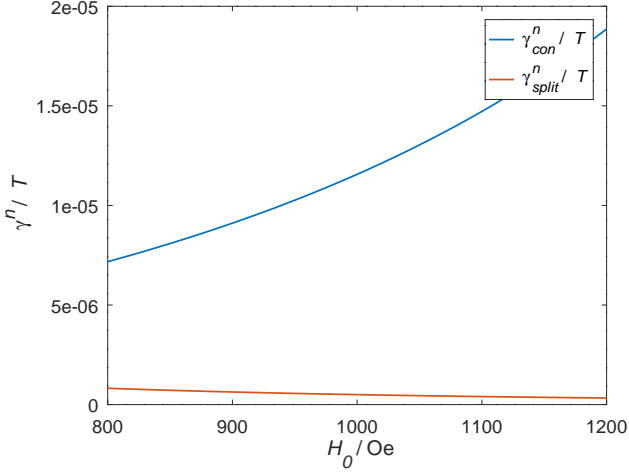}
 \caption{Magnetic field dependence of the magnon damping in thermal equilibrium
due to the confluence and the 
splitting processes. The plotted damping rates
$\gamma^n_{\mathrm{con}}$ and $\gamma^n_{\mathrm{split}}$ are obtained from
Eqs.~\eqref{eq:dampconf1} and \eqref{eq:dampsplit1}
by averaging  over all momenta $\kv$ satisfying $\epsilon_\kv=\omega_0$.
For the calculation we have assumed a film  thickness of $d=22.8\, \mu$m 
and a pumping frequency $\omega_0=13.857$ GHz.}
 \label{fig:damp_th}
\end{figure}

\begin{figure}[tb]
 \centering
 \includegraphics[clip=true,trim=10pt 180pt 40pt 200pt,width=\linewidth]{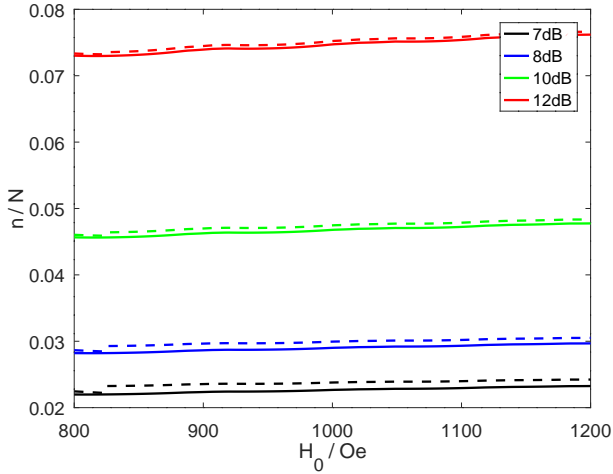}
 \caption{Magnetic field dependence of the stationry magnon density   
within S-theory given by Eq.~\eqref{eq:n_stat_S}
for the same parameters as in Fig.~\ref{fig:damp_th}.
The continuous lines are obtained  assuming a
constant magnon damping  $\gamma_\kv = 2.19 \times 10^{-3}$GHz, while the dashed lines 
are obtained  by substituting the equilibrium magnon damping 
shown in Fig.~\ref{fig:damp_th} into  Eq.~\eqref{eq:n_stat_S}.}
 \label{fig:md_th}
\end{figure}

Unfortunately, the approximations made in this section are only valid for small 
pumping energy $\left|V_\kv\right|$, whereas the experiment \cite{Noack19} has been performed
in the regime of parametric instability where $\left|V_\kv\right| > \left|E_\kv\right|$. Therefore we expect that the estimates for the magnon damping
in this subsection are not relevant for the experiment 
of Ref.~[\onlinecite{Noack19}].
This is also confirmed by the linear magnetic field dependence 
of the damping due to the confluent and the splitting processes 
in thermal equilibrium shown in Fig.~\ref{fig:damp_th},
which can be obtained by numerically evaluating Eqs.~\eqref{eq:magconf}
and \eqref{eq:magsplit}. 
In Fig.~\ref{fig:md_th} we show the corresponding
magnon density obtained by 
inserting this damping into the expression~\eqref{eq:n_stat_S} 
for the magnon distribution predicted by  S-theory.
Obviously, the magnetic-field dependence 
is linear in a wide range of fields and shows
a small discontinuity at $H_0\approx820$ Oe 
where the condition \eqref{eq:conf_quad} is violated.
By comparing Fig.~\ref{fig:md_th} with the
experimental result for the inverse spin-Hall voltage 
shown in Fig.~\ref{fig:experiment}, we conclude that by inserting the
equilibrium magnon damping into the S-theory result for the stationary
magnon density of the pumped magnon gas we cannot explain the experimental results.


\subsection{\label{sec:iterate_S}
Solution of the kinetic equations with microscopic collision integrals}

In this section we  show that the experimental results can be 
explained when the effect of collisions on the stationary distribution  of the pumped magnon gas is taken into account microscopically within a  
non-equilibrium many-body approach where we approximately solve 
the kinetic equations (\ref{eq:dgl_ci})
with collision integrals given in Appendix~C. As it stands, 
this system of non-linear integro-differential equations is very complicated and we have not been able to solve it directly. Fortunately, we have found an approximation strategy which
is sufficiently simple to allow for a numerical solution of the kinetic equations while it still 
contains the relevant physical processes which determine 
the detailed form of the experimentally observed 
inverse spin-Hall signal.
Our strategy is to divide the magnons into the following two groups corresponding to different regimes in momentum space and different energy windows:

\begin{enumerate}
 \item {\it{Parametric magnons}} are directly excited  by the
oscillating microwave field via parametric resonance.
From  S-theory\cite{Zakharov70,Zakharov74,Hahn20} we know  that only magnons in a small area of the momentum space near the resonance surface defined by $\epsilon_\kv=\omega_0/2$ are generated by the parametric pumping so that it is justified to assume that all parametric magnons fulfill the resonance condition $\epsilon_\kv=\omega_0/2$.

 \item {\it{Secondary magnons}} are created by confluence process of two 
parametric magnons. As a consequence, their energy $\epsilon_\kv=\omega_0$ is twice as large as the energy of parametric magnons.
\end{enumerate}
Assuming that the non-equilibrium magnon dynamics is dominated by these two groups of magnons,
we can approximate the distribution of all other magnons in the collision integrals by the
thermal equilibrium distribution.
These approximations significantly simplify the collision integrals as the arguments of the delta functions only vanish if two of the energies correspond to parametric magnons and the other one to secondary magnons. The complexity of evaluating the collision integrals 
numerically is  then greatly reduced. 
Neglecting the expectation values of the magnon operators we
find from the general expressions for the collision integrals
given in Appendix~C that
 the collision integrals associated with the two different magnon groups can be written as
\begin{widetext}
\begin{subequations}
\label{eq:col_int_groups}
\begin{eqnarray}
I^{n^{(1)}}_\kv &=& \frac{2\pi}{N} \sum_{\stackrel{\qv}{\stackrel{\epsilon_\qv=\omega_0}{\epsilon_\qkv=\omega_0/2}}} \biggl[
\left|\gam{\qv}{\kv}{\qkv}\right|^2 \left(n^{(2)}_\qv \left[1+n^{(1)}_\qkv\right] - n^{(1)}_\kv \left[n^{(1)}_\qkv - n^{(2)}_\qv\right]\right) 
 \nonumber
 \\
 & & \hspace{20mm}
+ \mathrm{Re}\left[\left(\gam{\qv}{\kv}{\qkv}\right)^*\gam{\qkv}{\kv}{\qv} (\tilde{p}^{(2)}_{\qv})^* \tilde{p}^{(1)}_\qkv\right]\biggr],
\\
I^{n^{(2)}}_\kv &=& \frac{2\pi}{N} \sum_{\stackrel{\qv}{\stackrel{\epsilon_\qv=\omega_0/2}{\epsilon_\kqv=\omega_0/2}}} \biggl[  \frac{1}{2}\left|\gam{\kv}{\qv}{\qkv}\right|^2 \left(n^{(1)}_\qv n^{(1)}_\kqv - n^{(2)}_\qv \left[1 + n^{(1)}_\qv + n^{(1)}_\kqv\right]\right) \nonumber\\
&& \hspace{20mm}   - \mathrm{Re}\left[ \gam{\kv}{\qv}{\kqv}\left(\gam{\qv}{\kqv}{\kv}\right)^* \tilde{p}^{(2)}_\kv (\tilde{p}^{(1)}_\qv)^* + \left(\gam{\kv}{\qv}{\qkv}\right)^*\gam{\kqv}{\kv}{\qv} \tilde{p}^{(2)}_\kv (\tilde{p}^{(1)}_\kqv)^*\right]  \biggr],\\
I^{p^{(1)}}_\kv &=& \frac{2\pi}{N} \sum_{\stackrel{\qv}{\stackrel{\epsilon_\qv=\omega_0}{\epsilon_\qkv=\omega_0/2}}} \biggl[
 \gam{\qkv}{\kv}{\qv}\left(\gam{\kv}{\qkv}{\qv}\right)^* \tilde{p}^{(1)}_\qkv \left(n^{(2)}_\qv-n^{(1)}_\kv\right) 
 \nonumber
 \\
 & & \hspace{18mm}
- \left|\gam{\qkv}{\kv}{\qv}\right|^2 \tilde{p}^{(1)}_\kv \left(1 + n^{(1)}_\qkv - \frac{1}{2}n^{(2)}_\qv - \tilde{p}^{(2)}_\qv\right)\biggr],
 \\
I^{p^{(2)}}_\kv &=& 0,
\end{eqnarray}
\end{subequations}
\end{widetext}
where $n^{(1)}_\kv$ and $\tilde{p}^{(1)}_\kv$ refer to the magnon distribution functions of parametric magnons and  $n^{(2)}_\kv$ and $\tilde{p}^{(2)}_\kv$ refer to the secondary magnon group.
When summing over the loop momentum $\qv$, 
we have to implement the  conditions $\epsilon_\kv=\epsilon_\qkv=\omega_0/2$, $\epsilon_\qv=\omega_0$ in the collision integrals of the parametric magnon group, 
and the conditions $\epsilon_\kv=\omega_0$ and $\epsilon_\qv=\epsilon_\kqv=\omega_0/2$ for the secondary magnon group. When all of these conditions can be 
fulfilled simultaneously, there is only one possible combination of wavevectors so that
only a single term contributes to the sums in Eq.~(\ref{eq:col_int_groups}).
In order to calculate the collision integrals numerically we thus have to find the 
specific combination of wavevectors that fulfill momentum and energy conservation. 
Then, we interpolate linearly between the magnon distribution functions defined on a finite grid in momentum space and evaluate the expressions \eqref{eq:col_int_groups}. 
It is also possible that for certain parameters the conservation laws cannot be 
fulfilled, so that the collision integrals vanish in our approximation.
All other magnons which do not belong to the above two groups 
are assumed to be in thermal equilibrium where the stationary distributions are given by the Bose-Einstein distribution 
(\ref {eq:BoseEinstein}) 
with $T = 290 $ K. We take the contribution of
these equilibrium magnons to the damping of the non-equilibrium
magnons  into account using the equilibrium damping rates 
derived in Sec.~\ref{sec:dampeq}.

To obtain a self-consistent solution of the kinetic equations (\ref{eq:dgl_ci}) with collision integrals given by Eq.~(\ref{eq:col_int_groups}) we use the following iterative procedure:
Initially,  we completely neglect the collision integrals and use
the stationary distribution~(\ref{eq:n_stat}) of the kinetic equations
with phenomenological damping
$\gamma^n_\kv=\gamma^p_\kv=\gamma_0=2.87 \times 10^{-3}$GHz 
to construct the initial seed for the iteration. 
We then substitute  the resulting stationary distribution back into our microscopic
expressions (\ref{eq:col_int_groups}) for the collision integrals
and calculate a new estimate for the collision integrals. Next, we
use the result to re-calculate a refined estimate for the stationary solution of 
the kinetic equations~\eqref{eq:n_stat}.
To obtain new values for non-equilibrium damping rates 
$\gamma^{n^{(1)}}_\kv$ and $\gamma^{p^{(1)}}_\kv$ 
we assume that the terms proportional to $n^{(1)}_\kv$ and $\tilde{p}^{(1)}_\kv$ 
dominate the collision integrals 
and estimate $\gamma^{n^{(1)}}_\kv$ and $\gamma^{p^{(1)}}_\kv$ by 
$I^{n^{(1)}}_\kv / n^{(1)}_\kv$ and $I^{p^{(1)}}_\kv / \tilde{p}^{(1)}_\kv$.
The result is again substituted into the
right-hand side of the collision integrals (\ref{eq:col_int_groups}) and the procedure
is iterated again.
Gradually, we obtain corrections to the initial estimate of the magnon distribution
in the stationary non-equilibrium state.
To control the convergence of this algorithm we estimate the error by evaluating the derivatives $\partial_t n_\kv$ and $\partial_t \tilde{p}_\kv$ given by the equations of motion \eqref{eq:dgl_ci} and summing up the absolute values for every magnon mode. This expression should vanish if our  estimates for the magnon distributions  approach the exact stationary 
solutions. 
If this estimated error tends to zero during the iteration, 
our algorithm has produced a self-consistent stationary solution of the
kinetic equations (\ref{eq:dgl_ci}).
Note that 
the  vanishing 
of the off-diagonal collision integral $I^{p^{(2)}}_\kv$
associated with the secondary magnons implies that the stationary solution
of the kinetic equation~\eqref{eq:n_stat_cci} has the property that $n^{(2)}_\kv$ vanishes  independently of the value of $\tilde{p}^{(2)}_\kv$.

\begin{figure}[tb]
 \centering
 \includegraphics[clip=true,trim=20pt 180pt 40pt 200pt,width=\linewidth]{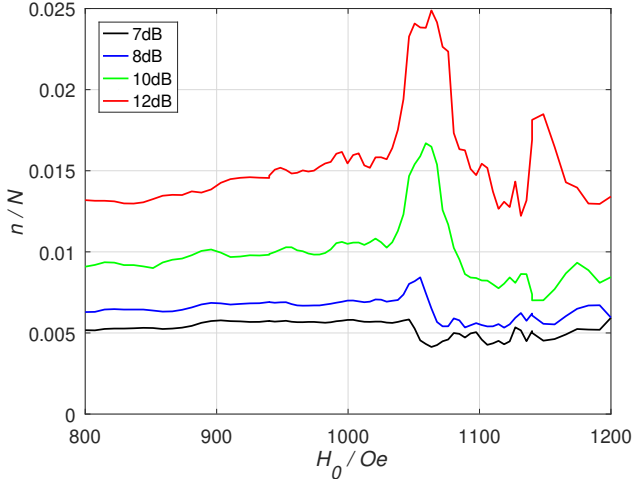}
 \caption{The magnon density obtained by the procedure described in Sec.~\ref{sec:iterate_S} for a thin YIG film of thickness $d=22.8 \mu$m and $\omega_0=13.857$ GHz is plotted over the external field strength $H_0$ for four  different pumping strengths. The parameter for the pumping strength is $\langle V_{\kv_1}-\gamma_0 \rangle_{\kv_1}$ with the average taken over all momenta $\kv_1$ of parametric magnons. Our theoretical result shown in this figure should be compared with the experimental results by Noack {\it{et al.}}\cite{Noack19} reproduced in Fig.~\ref{fig:experiment}.
}
 \label{fig:n_stat_it_S}
\end{figure}

In Fig.~\ref{fig:n_stat_it_S} we show our numerical results
for a YIG film with thickness $d=22.8\mu$m 
(corresponding to $N = 18 423$) in a
microwave field with frequency $\omega_0=13.857$GHz 
for four different pumping strengths between 7dB and 12dB, where the parameter controlling the pumping strength is $\langle V_{\kv_1}-\gamma_0 \rangle_{\kv_1}$ with the average taken over all momenta $\kv_1$ of parametric magnons.
The magnon density shown in Fig.~\ref{fig:n_stat_it_S} is approximated by taking the sum over all magnon modes used for the calculations,
\begin{equation}
 n^s = \sum_{i=1}^{N_\theta} n^s_i,
\end{equation}
where the momentum dependence of the 
magnon distribution functions are parameterized by the 
angle $\theta_i = \theta_{\bd{k}_i} $ of the in-plane wavevectors defined in Eq.~(\ref{eq:inplanewave})
and we use $N_\theta=40$ angles
of equal size in the interval $[0,\pi/2]$.
The wavevectors $\kv_1$ and $\kv_2$ of parametric and secondary magnons for a given angle $\theta_i$ are calculated by solving the equations $\epsilon_{\kv_1}=\omega_0/2$ and $\epsilon_{\kv_2}=\omega_0$ numerically for $\kv_1$ and $\kv_2$ 
with magnon dispersion given by Eq.~\eqref{eq:disp_rel}.
%
%
Apart from a small offset in the overall field strength by about $50$ Oe,
the main features of the experimentally observed line-shape of the
inverse spin-Hall signal shown in Fig.~\ref{fig:experiment}
are reproduced remarkably well by our calculation.
Recall that S-theory with phenomenological constant damping cannot
explain this line-shape.
In particular, the experimentally observed dip around $H_0 \approx 1050$ Oe for small pumping strength which evolves into a peak at the same field 
for larger pumping strength is  reproduced by our method.
Note, however, that  in the experiment these features appear at a slightly lower field of 
$H_0 \approx 1000$ Oe. A possible explanation for this
discrepancy in the overall field strength is 
the influence of cubic crystallographic and uni-axial anisotropy fields which can 
modify the saturation magnetization.  
It is therefore plausible that
the experimentally relevant value of the
saturation magnetization  differs from the value of $ 1750$ G assumed 
in our calculation which can explain the $50$ Oe shift in the
position of the peaks and dips  in the upper and lower part of Fig.~\ref{fig:n_stat_it_S} .

\begin{figure}[tbhp]
\begin{minipage}{10pt}
(a)
\end{minipage}
\begin{minipage}{0.92\linewidth}
\includegraphics[clip=true,trim=20pt 180pt 45pt 180pt,width=\linewidth]{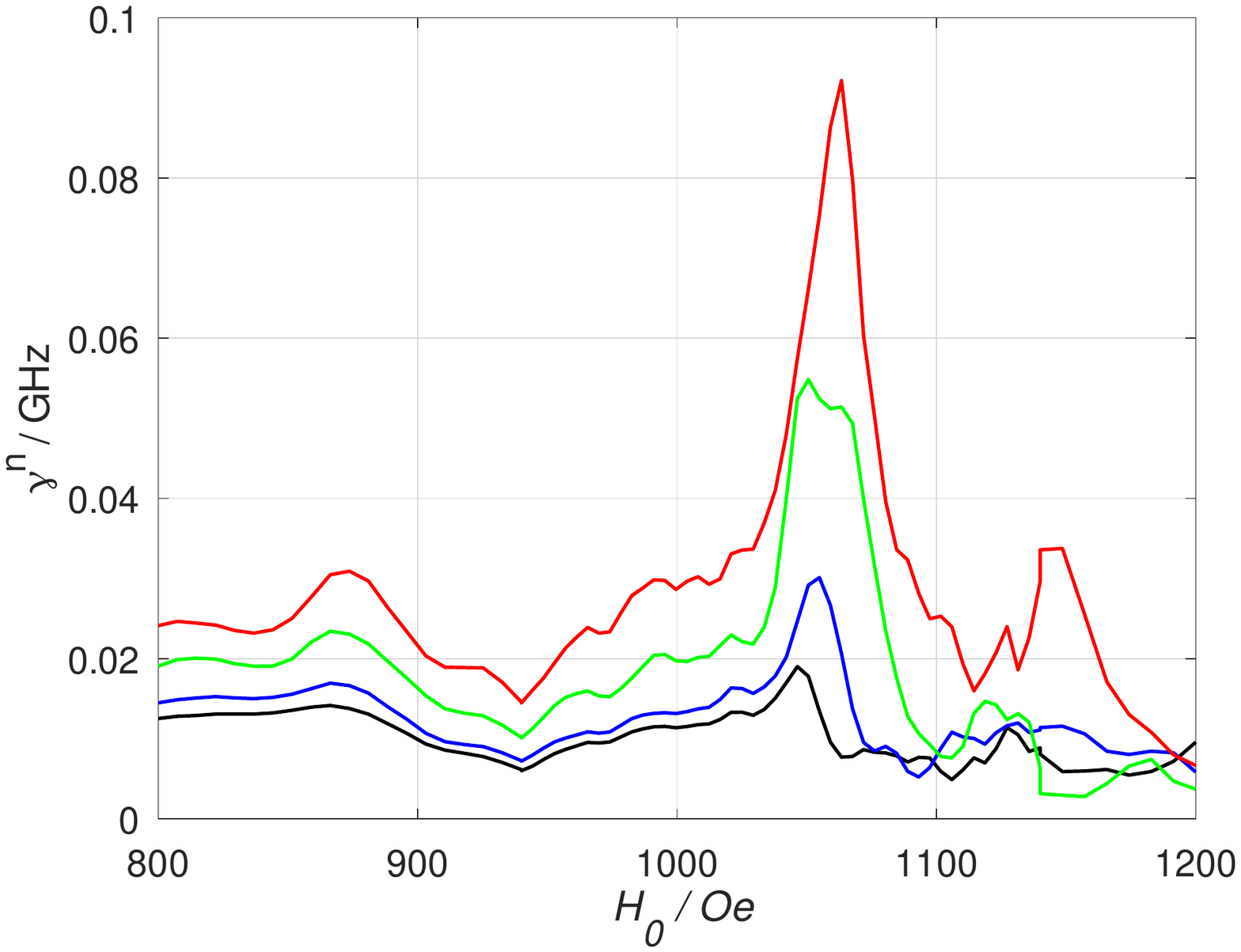}
\end{minipage}
\begin{minipage}{10pt}
(b)
\end{minipage}
\begin{minipage}{0.92\linewidth}
\includegraphics[clip=true,trim=20pt 180pt 45pt 180pt,width=\linewidth]{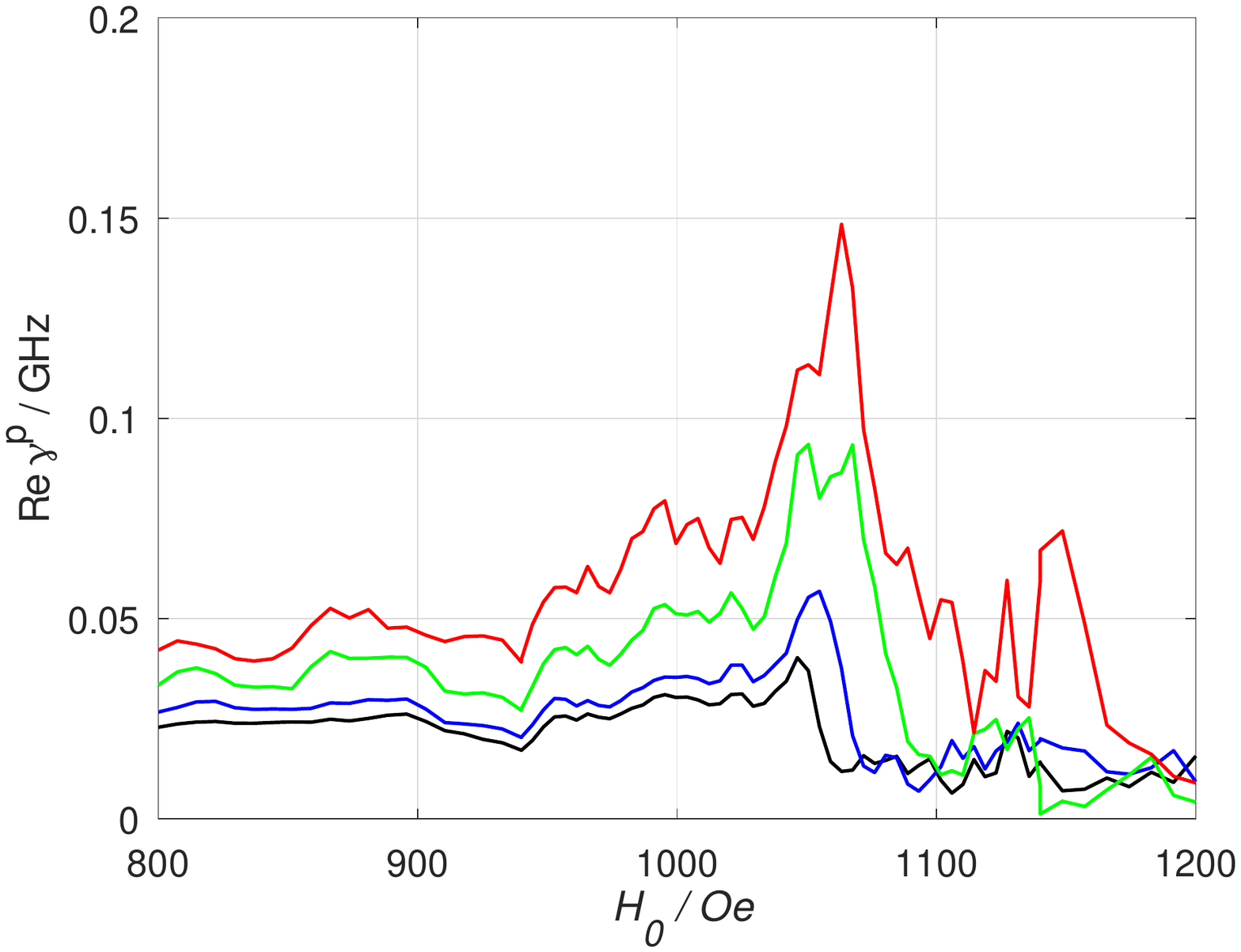}
\end{minipage}
\caption{The damping defined by Eqs.~\eqref{eq:gamma_approx} in the stationary non-equilibrium state shown in Fig.~\ref{fig:n_stat_it_S} is plotted over the external field strength $H_0$ for the same parameter values as in Fig.~\ref{fig:n_stat_it_S}.
}
\label{fig:damp}
\end{figure}

To show that dip and the peak  are related to the confluent magnon damping,
we have plotted in Fig.~\ref{fig:damp} the cumulative damping rates
$\gamma^n=\sum_{i=1}^{N_\theta} \gamma^n_i$ and 
${\rm Re} \gamma^p=\sum_{i=1}^{N_\theta} {\rm Re} \gamma^p_i$ for the stationary non-equilibrium state we have obtained from our kinetic equations.
Obviously, the peaks in the cumulative magnon damping are
observed at the same magnetic field strength where the enhancement of the magnon density takes place.
Not all magnon modes show these enhancements. The distribution functions for most of the magnon modes still increase linearly with the external field strength and 
only a few  magnon modes around $\theta_\kv\approx 40^{\circ}$ have peaks between $H_0=1050$~Oe and $H_0=1100$~Oe.

It is interesting to compare the order of magnitude of the 
cumulative non-equilibrium damping $\gamma^n$
shown in the upper panel of Fig.~\ref{fig:damp} with the established value
of the Gilbert damping used in phenomenological approaches for YIG~\cite{Bender14,Hoffman13,Cornelissen16}. Usually  the momentum-dependent damping
$\gamma_{\bd{k}} $ is parameterized in terms of a dimensionless damping parameter
$\alpha = \gamma_{\kv}/ ( 2 \epsilon_\kv )$, where $\epsilon_\kv$ is the magnon dispersion.\cite{Bender14} 
According to Refs.~[\onlinecite{Hoffman13,Cornelissen16}]   for thermal acoustic magnons in
YIG the typical value of
$\alpha$ is for small wavevectors  of order $10^{-4}$.
On the other hand, our cumulative non-equilibrium damping 
$\gamma^n$ in the upper panel of Fig.~\ref{fig:damp} is typically of order
$0.02$ GHz,
which yields a dimensionless damping parameter $\alpha
 \approx 1.4 \times 10^{-3}$. We conclude that the non-equilibrium damping
obtained within our microscopic approach is roughly an order of magnitude larger
than the accepted phenomenological value of the equilibrium damping of thermal magnons in YIG.

The rather complicated dependence of the non-equilibrium
magnon density on the external magnetic field shown in Fig.~\ref{fig:damp})
cannot be reproduced within conventional S-theory where the microscopic collision integrals 
are replaced by a phenomenological relaxation rate. 
In the relevant parameter regime, S-theory predicts a linear 
dependence of the magnon density on the external field strength as shown in Fig.~\ref{fig:mdS}.
Note also that within S-theory the damping is assumed to be strong
so that only magnon modes near the maximum of the pumping energy $V_\kv$ at $\theta_\kv=90^{\circ}$ are significantly occupied. In fact, the magnon modes which we have 
identified to be responsible for the observed peaks and dips 
are assumed to be suppressed by the phenomenological damping in S-theory.
Thus, it is evident that the experimentally observed structures 
in the non-equilibrium magnon density
are caused by the confluence and splitting decay
processes; the kinematic constraints controlling these 
processes are fully taken into account
in our collision integrals which 
couple pairs of parametric magnons at special 
wavevectors depending on the external field strength.
The mathematical structure of the equations of motion is complicated and leads to peak structures appearing in the collision integrals at certain field strengths. This in turn 
gives rise to  similar structures  in the field-dependent magnon density 
close to magnetic fields
where confluent magnon decay is kinematically possible.

\section{\label{sec:conclusions}
Summary and conclusions}

In this work we have derived and solved kinetic equations  for   
pumped magnons  in YIG with  collision integrals  discribing
dissipative effects associated with magnon decays.
The collisionless limit  of these
equations has recently been discussed in Ref.~[\onlinecite{Hahn20}]. 
However, to explain
recent experimentel data \cite{Noack19} for the magnetic field dependence of the 
inverse spin-Hall voltage in the stationary non-equilibrium state
of pumped magnons in YIG
a microscopic understanding of magnon decays is crucial.
We have derived the relevant  
collision integrals due to cubic interaction vertices 
using a systematic  expansion in powers of connected equal-time 
correlations \cite{Fricke97}.
We have obtained the collision integrals for the diagonal and off-diagonal distribution functions containing terms which are linear and  quadratic in the magnon distribution functions 
as well as the  expectation values of the magnon operators.
In previous works these collision integrals were not taken into account due to their complexity or were only derived within Born approximation \cite{Vinikovetskii79}  and 
evaluated in thermal equilibrium.

We have found a way to numerically solve the resulting kinetic equations 
within an approximation where only two groups of magnons are asumed to be
driven out of equilibrium: 
parametric magnons that are generated by the pumping,  
and secondary magnons that are involved in confluence and splitting processes described by the microscopic collision integrals. We have explicitly constructed
the stationary non-equilibrium solution of the kinetic equations for the pumped magnon gas. 

Our results show in a large parameter regime
a roughly linear magnetic field dependence
of the magnon density, in agreement with previous results obtained within 
a collisionless kinetic theory. However, near the magnetic field strength 
where magnon decays (confluence and splitting processes) 
become kinematically allowed, we have obtained 
peak and dip structures in the magnon density, in good agreement with the
experiment by Noack {\it{et al.}} \cite{Noack19} where the non-equilibrium magnon density 
has been  measured via the inverse spin-Hall effect. 

\section*{ACKNOWLEDGEMENTS}
We are grateful to  A. A. Serga for his comments on the manuscript. We also thank  A. A. Serga and 
T. Noack for helping us to prepare Fig.~\ref{fig:experiment} and for their permission 
to present the experimental data of 
Ref.~[\onlinecite{Noack19}] in this figure.

\appendix
\setcounter{equation}{0}
\renewcommand{\theequation}{A\arabic{equation}}
\section*{\label{sec:ward}
APPENDIX A: HAMILTONIAN FOR PUMPED MAGNONS IN YIG}
Here we derive the 
magnon Hamiltonian for the parametrically pumped magnon gas in YIG 
following mainly Ref.~[\onlinecite{Kreisel09}]. We start from the effective spin Hamiltonian 
for YIG~\cite{Cherepanov93,Lvov94,Rezende09,Kloss10,Hick10,Rueckriegel14,Rezende06,Kreisel09} given in Eq.~(\ref{eq:H_spin}).
The exchange couplings $J_{ij}$ assume the value  $J \approx 1.29$ K for all pairs of nearest neighbor spins located at lattices sites $\bd{r}_i$ and $\bd{r}_j$,  
and the dipolar tensor is~\cite{Kreisel09,Filho00}
\begin{equation}
D_{i j}^{\alpha \beta} = \left(1-\delta_{i j}\right)\frac{\mu^2}{\left|\bd{r}_{i j}\right|^3}\left[3\hat{{r}}_{i j}^\alpha\hat{{r}}_{i j}^\beta-\delta^{\alpha\beta}\right] ,
 \label{eq:Ddef}
\end{equation}
where $\mu$ is the magnetic moment of the spins,
$\bd{r}_{ij}=\bd{r}_i-\bd{r}_j$,  and $\hat{\bd{r}}_{ij}=\bd{r}_{ij}/\left|\bd{r}_{ij}\right|$.
After Holstein-Primakoff transformation \cite{Holstein40} and expansion in powers of
$1/S$  the spin Hamiltonian is mapped onto an effective boson Hamiltonian of the form
\eqref{eq:H_1/S} where the terms ${\cal{H}}_i$  can be expressed  in terms of
Holstein-Primakoff bosons $b_i$ and $b^\dagger_i$.
The zeroth order contribution $\mathcal{H}_0(t)$ can be 
dropped as it does not contain any boson operators.
Transforming   to momentum space, 
\begin{equation}
b_i = \frac{1}{\sqrt{N}} \sum\limits_\kv \text{e}^{i \kv \cdot \bd{r}_i} b_\kv,
\end{equation}
where $N$ is the total number of lattice sites,
the contributions to the Hamiltonian up to fourth order in the bosons can be written as~\cite{Hick10}
\begin{subequations}
\begin{eqnarray}
\mathcal{H}_2(t)&=&\sum\limits_\kv\left[A_\kv b^\dagger_\kv b_\kv + \frac{B_\kv}{2}\left(b^\dagger_\kv b^\dagger_{-\kv}+b_{-\kv}b_\kv\right)\right] \nonumber\\*
&&\hspace{8mm} + h_1\cos\left(\omega_0t\right)\sum\limits_\kv b^\dagger_\kv b_\kv ,
\label{eq:H2_ft}\\
\mathcal{H}_3 &=& \frac{1}{\sqrt{N}} \sum\limits_{\kv_1, \kv_2, \kv_3} \delta_{\kv_1+\kv_2+\kv_3, 0} \frac{1}{2!} \left[\Gamma^{\bar{b}bb}_{1; 2, 3} b^\dagger_{-1} b_2 b_3 \right.\nonumber\\
&&\hspace{16mm}\left.+ \Gamma^{\bar{b}\bar{b}b}_{1, 2; 3} b^\dagger_{-1} b^\dagger_{-2} b_3\right] ,\\
\mathcal{H}_4 &=& \frac{1}{N} \sum\limits_{\kv_1, \dots, \kv_4} \delta_{\kv_1+\dots+\kv_4, 0} \left[\frac{1}{\left(2!\right)^2} \Gamma^{\bar{b}\bar{b}bb}_{1, 2; 3, 4} b^\dagger_{-1} b^\dagger_{-2} b_3 b_4 \right.\nonumber\\*
&&\hspace{-2mm}\left.+ \frac{1}{3!} \Gamma^{\bar{b}bbb}_{1; 2, 3, 4} b^\dagger_{-1} b_2 b_3 b_4  + \frac{1}{3!} \Gamma^{\bar{b}\bar{b}\bar{b}b}_{1, 2, 3; 4} b^\dagger_{-1} b^\dagger_{-2} b^\dagger_{-3} b_4\right]. \nonumber\\\label{eq:H4_ft}
\end{eqnarray}
\end{subequations}
The vertices in \eqref{eq:H2_ft}-\eqref{eq:H4_ft}  
can be expressed in terms of the Fourier transforms of the 
exchange and dipolar couplings,
\begin{subequations}
\begin{eqnarray}
J_\kv &=& \sum\limits_i \text{e}^{-i \kv \cdot \bd{r}_{i j}} J_{i j}, \\*
D^{\alpha \beta}_\kv &=& \sum\limits_i \text{e}^{-i \kv \cdot \bd{r}_{i j}} D^{\alpha \beta}_{i j}.
\end{eqnarray}
\end{subequations}
The coefficients $A_\kv$ and $B_\kv$ in Eq.\eqref{eq:H2_ft} are
\begin{subequations}
\begin{eqnarray}
A_\kv &=& h_0 + S \left(J_{\bd{0}} - J_\kv\right) + S \left[D^{zz}_{\bd{0}} - \frac{1}{2}\left(D^{xx}_\kv + D^{yy}_\kv\right)\right], \nonumber\\
\label{eq:Akdef} \\
B_\kv &=& -\frac{S}{2} \left[D^{xx}_\kv - 2 i D^{xy}_\kv - D^{yy}_\kv\right],
\label{eq:Bkdef}
\end{eqnarray}
\end{subequations}
while the cubic vertices depend only on the dipolar tensor as follows,
\begin{subequations}
\begin{eqnarray}
\Gamma^{\bar{b}bb}_{1; 2, 3} &=& \sqrt{\frac{S}{2}} \left[D^{zy}_{\kv_2} - i D^{zx}_{\kv_2} + D^{zy}_{\kv_3} - i D^{zx}_{\kv_3} \right.\nonumber\\
&&\left.+ \frac{1}{2} \left(D^{zy}_{\mathbf{0}} - i D^{zx}_{\mathbf{0}}\right)\right] , \\
\Gamma^{\bar{b}\bar{b}b}_{1, 2; 3} &=& \left(\Gamma^{\bar{b}bb}_{3; 2, 1}\right)^* ,
\end{eqnarray}
 \end{subequations}
and the quartic vertices are
 \begin{subequations}
 \begin{eqnarray}
\Gamma^{\bar{b}\bar{b}bb}_{1, 2; 3, 4} &=& -\frac{1}{2}\left[J_{\kv_1+\kv_3} + J_{\kv_2+\kv_3} + J_{\kv_1+\kv_4} + J_{\kv_2+\kv_4} \right.\nonumber\\*
&&\left.+ D^{zz}_{\kv_1+\kv_3} + D^{zz}_{\kv_2+\kv_3} + D^{zz}_{\kv_1+\kv_4} + D^{zz}_{\kv_2+\kv_4}\right.\nonumber\\*
&&\left. - \sum\limits_{i=1}^4 \left(J_{\kv_i} - 2 D^{zz}_{\kv_i}\right)\right], \\
\Gamma^{\bar{b}bbb}_{1; 2, 3, 4} &=& \frac{1}{4} \left[D^{xx}_{\kv_2} - 2 i D^{xy}_{\kv_2} - D^{yy}_{\kv_2} + D^{xx}_{\kv_3}-2 i D^{xy}_{\kv_3} - D^{yy}_{\kv_3} \right.\nonumber\\*
&&\left. + D^{xx}_{\kv_4} - 2 i D^{xy}_{\kv_4} - D^{yy}_{\kv_4}\right], \\*
\Gamma^{\bar{b}\bar{b}\bar{b}b}_{1, 2, 3; 4} &=& \left(\Gamma^{\bar{b}bbb}_{4; 1, 2, 3}\right)^* .
\end{eqnarray}
\end{subequations}
Next, we diagonalize the 
time-independent part of $\mathcal{H}_2(t)$ by introducing 
magnon annihilation and creation operators $a_{\bd{k}}$ and $a^{\dagger}_{\bd{k}}$ 
via the Bogoliubov transformation,
\begin{equation}
\left( \begin{array}{c} b_{\bd{k}} \\ b^{\dagger}_{ - \bd{k}} \end{array} \right)
= \left( \begin{array}{cc} u_{\bd{k}} & - v_{\bd{k}} \\
- v_{\bf{k}}^{\ast} & u_{\bd{k}} \end{array} \right)
\left( \begin{array}{c} a_{\bd{k}} \\ a^{\dagger}_{ - \bd{k}} \end{array} \right),
\end{equation}
where
\begin{subequations}
\begin{eqnarray}
u_\kv &=& \sqrt{\frac{A_\kv + \varepsilon_\kv}{2 \varepsilon_\kv}}, \\
v_\kv &=& \frac{B_\kv}{|B_\kv|} \sqrt{\frac{A_\kv - \varepsilon_\kv}{2 \varepsilon_\kv}},\\
\varepsilon_\kv &=& \sqrt{A_\kv^2 - |B_\kv|^2} .
 \label{eq:magdisp}
\end{eqnarray}
\end{subequations}
In terms of the magnon operators the time-dependent term in Eq.~\eqref{eq:H2_ft} 
leads to off-diagonal terms, so that the total quadratic  Hamiltonian reads~\cite{Hick10},
\begin{eqnarray}
\mathcal{H}_2(t) &=& \sum\limits_\kv \left[\varepsilon_\kv a^\dagger_\kv a_\kv + \frac{\varepsilon_\kv - A_\kv}{2} \right.\nonumber\\*
&&\left.\hspace{-14pt}+ h_1 \cos\left(\omega_0 t\right) \left(\frac{A_\kv}{\varepsilon_\kv} a^\dagger_\kv a_\kv - \frac{\varepsilon_\kv - A_\kv}{2 \varepsilon_\kv}\right)\right] \nonumber\\*
&&\hspace{-14pt}+ \sum\limits_\kv \left[V_\kv \cos\left(\omega_0 t\right) a^\dagger_\kv a^\dagger_{-\kv} + V_\kv^* \cos\left(\omega_0 t\right) a_{-\kv}a_\kv\right], \nonumber\\*
\label{eq:H2_r}
\end{eqnarray}
with pumping energy
\begin{equation}
V_\kv = -\frac{h_1 B_\kv}{2 \varepsilon_\kv} .
 \label{eq:magV}
\end{equation}
Expressing also the cubic and quartic parts of the Hamiltonian in terms of
magnon operators we obtain~\cite{Hick10}
\begin{widetext}
\begin{eqnarray}
\mathcal{H}_3 &=& \frac{1}{\sqrt{N}} \sum\limits_{\kv_1, \kv_2, \kv_3} \delta_{\kv_1+\kv_2+\kv_3, 0} \left[\frac{1}{2} \Gamma^{\bar{a}aa}_{1; 2, 3} a^\dagger_{-1} a_2 a_3 + \frac{1}{2} \Gamma^{\bar{a}\bar{a}a}_{1, 2; 3} a^\dagger_{-1} a^\dagger_{-2} a_3 + \frac{1}{3!} \Gamma^{aaa}_{1, 2, 3} a_1 a_2 a_3 + \frac{1}{3!} \Gamma^{\bar{a}\bar{a}\bar{a}}_{1, 2, 3} a^\dagger_{-1} a^\dagger_{-2} a^\dagger_{-3}\right],\nonumber\\*\\
\mathcal{H}_4 &=& \frac{1}{N} \sum\limits_{\kv_1, \kv_2, \kv_3, \kv_4} \delta_{\kv_1+\kv_2+\kv_3+\kv_4, 0} \left[\frac{1}{\left(2!\right)^2}\Gamma^{\bar{a}\bar{a}aa}_{1, 2; 3, 4} a^\dagger_{-1} a^\dagger_{-2} a_3 a_4 + \frac{1}{3!} \Gamma^{\bar{a}aaa}_{1; 2, 3, 4} a^\dagger_{-1} a_2 a_3 a_4 \right.\nonumber\\*
&&\left. + \frac{1}{3!} \Gamma^{\bar{a}\bar{a}\bar{a}a}_{1, 2, 3; 4} a^\dagger_{-1} a^\dagger_{-2} a^\dagger_{-3} a_4 + \frac{1}{4!} \Gamma^{aaaa}_{1, 2, 3, 4} a_1 a_2 a_3 a_4 + \frac{1}{4!} \Gamma^{\bar{a}\bar{a}\bar{a}\bar{a}}_{1, 2, 3, 4} a^\dagger_{-1} a^\dagger_{-2} a^\dagger_{-3} a^\dagger_{-4} \right] ,
\end{eqnarray}
with cubic vertices given by
\begin{subequations}
\begin{eqnarray}
\Gamma^{aaa}_{1, 2, 3} &=& - \Gamma^{\bar{b}bb}_{1; 2, 3} v_1 u_2 u_3 - \Gamma^{\bar{b}bb}_{2; 1, 3} v_2 u_1 u_3 - \Gamma^{\bar{b}bb}_{3; 1, 2} v_3 u_1 u_3 + \Gamma^{\bar{b}\bar{b}b}_{1, 2; 3} v_1 v_2 u_3 + \Gamma^{\bar{b}\bar{b}b}_{2, 3; 1} v_2 v_3 u_1 + \Gamma^{\bar{b}\bar{b}b}_{1, 3; 2} v_1 v_3 u_2 , \\
\Gamma^{\bar{a}aa}_{1; 2, 3} &=& \Gamma^{\bar{b}bb}_{1; 2, 3} u_1 u_2 u_3 + \Gamma^{\bar{b}bb}_{2; 1, 3} v_1 v_2 u_3 + \Gamma^{\bar{b}bb}_{3; 1, 2} v_1 v_3 u_2 - \Gamma^{\bar{b}\bar{b}b}_{3, 2; 1} v_3 v_2 v_1 - \Gamma^{\bar{b}\bar{b}b}_{1, 2; 3} v_2 u_1 u_3 - \Gamma^{\bar{b}\bar{b}b}_{1, 3; 2} v_3 u_1 u_2 , \\
\Gamma^{\bar{a}\bar{a}a}_{1, 2; 3} &=& \left(\Gamma^{\bar{a}aa}_{3; 2, 1}\right)^* , \\
\Gamma^{\bar{a}\bar{a}\bar{a}}_{1, 2, 3} &=& \left(\Gamma^{aaa}_{1, 2, 3}\right)^* ,
\end{eqnarray}
\label{eq:cubic_vertices}
\end{subequations}
and quartic vertices
\begin{subequations}
\begin{eqnarray}
\Gamma^{aaaa}_{1, 2, 3, 4} &=& \Gamma^{\bar{b}\bar{b}bb}_{1, 2; 3, 4} u_1 u_2 v_3 v_4 + \Gamma^{\bar{b}\bar{b}bb}_{1, 3; 2, 4} u_1 u_3 v_2 v_4 + \Gamma^{\bar{b}\bar{b}bb}_{1, 4; 2, 3} u_1 u_4 v_2 v_3 + \Gamma^{\bar{b}\bar{b}bb}_{2, 3; 1, 4} u_2 u_3 v_1 v_4 \nonumber\\*
&& + \Gamma^{\bar{b}\bar{b}bb}_{2, 4; 1, 3} u_2 u_4 v_1 v_3 + \Gamma^{\bar{b}\bar{b}bb}_{3, 4; 1, 2} u_3 u_4 v_1 v_2 \nonumber\\*
&& - \Gamma^{\bar{b}bbb}_{4; 1, 2, 3} u_1 u_2 u_3 v_4 - \Gamma^{\bar{b}bbb}_{3; 1, 2, 4} u_1 u_2 u_4 v_3 - \Gamma^{\bar{b}bbb}_{2; 1, 3, 4} u_1 u_3 u_4 v_2 - \Gamma^{\bar{b}bbb}_{1; 2, 3, 4} u_2 u_3 u_4 v_1 \nonumber\\*
&& - \Gamma^{\bar{b}\bar{b}\bar{b}b}_{2, 3, 4; 1} u_1 v_2 v_3 v_4 - \Gamma^{\bar{b}\bar{b}\bar{b}b}_{1, 3, 4; 2} u_2 v_1 v_3 v_4 - \Gamma^{\bar{b}\bar{b}\bar{b}b}_{1, 2, 4; 3} u_3 v_1 v_2 v_4 - \Gamma^{\bar{b}\bar{b}\bar{b}b}_{1, 2, 3; 4} u_4 v_1 v_2 v_3,\\
\Gamma^{\bar{a}aaa}_{1; 2, 3, 4} &=& -\Gamma^{\bar{b}\bar{b}bb}_{2, 1; 3, 4} u_2 v_1 v_3 v_4 - \Gamma^{\bar{b}\bar{b}bb}_{3, 1; 2, 4} u_3 v_1 v_2 v_4 - \Gamma^{\bar{b}\bar{b}bb}_{4, 1; 2, 3} u_4 v_1 v_2 v_3 - \Gamma^{\bar{b}\bar{b}bb}_{2, 3; 1, 4} u_2 u_3 u_1 v_4 \nonumber\\*
&& - \Gamma^{\bar{b}\bar{b}bb}_{2, 4; 1, 3} u_2 u_4 u_1 v_3 - \Gamma^{\bar{b}\bar{b}bb}_{3, 4; 1, 2} u_3 u_4 u_1 v_2 \nonumber\\*
&& + \Gamma^{\bar{b}bbb}_{1; 2, 3, 4} u_1 u_2 u_3 u_4 + \Gamma^{\bar{b}bbb}_{4; 3, 2, 1} u_3 u_2 v_1 v_4 + \Gamma^{\bar{b}bbb}_{3; 4, 2, 1} u_4 u_2 v_1 v_3 + \Gamma^{\bar{b}bbb}_{2; 4, 3, 1} u_4 u_3 v_1 v_2 \nonumber\\*
&& + \Gamma^{\bar{b}\bar{b}\bar{b}b}_{1, 2, 3; 4} u_4 u_1 v_2 v_3 + \Gamma^{\bar{b}\bar{b}\bar{b}b}_{1, 2, 4; 3} u_3 u_1 v_2 v_4 + \Gamma^{\bar{b}\bar{b}\bar{b}b}_{1, 3, 4; 2} u_2 u_1 v_3 v_4 + \Gamma^{\bar{b}\bar{b}\bar{b}b}_{4, 3, 2; 1} v_4 v_2 v_3 v_1 , \\
\Gamma^{\bar{a}\bar{a}aa}_{1, 2; 3, 4} &=& \Gamma^{\bar{b}\bar{b}bb}_{1, 2; 3, 4} u_1 u_2 u_3 u_4 + \Gamma^{\bar{b}\bar{b}bb}_{1, 3; 4, 2} u_1 u_4 v_3 v_2 + \Gamma^{\bar{b}\bar{b}bb}_{1, 4; 3, 2} u_1 u_3 v_4 v_2 + \Gamma^{\bar{b}\bar{b}bb}_{2, 3; 4, 1} u_2 u_4 v_3 v_1 \nonumber\\*
&& + \Gamma^{\bar{b}\bar{b}bb}_{2, 4; 3, 1} u_2 u_3 v_4 v_1 + \Gamma^{\bar{b}\bar{b}bb}_{3, 4; 2, 1} v_1 v_2 v_3 v_4 \nonumber\\*
&& - \Gamma^{\bar{b}bbb}_{4; 3, 2, 1} u_3 v_2 v_1 v_4 - \Gamma^{\bar{b}bbb}_{3; 4, 2, 1} u_4 v_2 v_1 v_3 - \Gamma^{\bar{b}bbb}_{2; 3, 4, 1} u_2 u_3 u_4 v_1 - \Gamma^{\bar{b}bbb}_{1; 3, 4, 2} u_1 u_3 u_4 v_2 \nonumber\\*
&& - \Gamma^{\bar{b}\bar{b}\bar{b}b}_{2, 3, 4; 1} u_2 v_3 v_4 v_1 - \Gamma^{\bar{b}\bar{b}\bar{b}b}_{1, 3, 4; 2} u_1 v_3 v_4 v_2 - \Gamma^{\bar{b}\bar{b}\bar{b}b}_{1, 2, 4; 3} u_1 u_2 u_3 v_4 - \Gamma^{\bar{b}\bar{b}\bar{b}b}_{1, 2, 3; 4} u_1 u_2 u_4 v_3 , \\
\Gamma^{\bar{a}\bar{a}\bar{a}\bar{a}}_{1, 2, 3, 4} &=& \Gamma^{aaaa}_{1, 2, 3, 4}, \\
\Gamma^{\bar{a}\bar{a}\bar{a}a}_{1, 2, 3; 4} &=& \left(\Gamma^{\bar{a}aaa}_{4; 3, 2, 1}\right)^*.
\end{eqnarray}
\label{eq:quartic_vertices}
\end{subequations}
\end{widetext}
Finally, let us give simplified expressions
for the Fourier transforms $J_\kv$ and $D^{\alpha\beta}_\kv$ for the geometry shown
in Fig.~\ref{fig:setup}
which reduce the complexity of
the coefficients $A_\kv$ and $B_\kv$ and the higher-order vertices.
For the energy scales probed in the experiment\cite{Noack19}
it is sufficient to retain only the lowest magnon band, so that we can derive 
the dispersion from an effective in-plane Hamiltonian. The simplest approximation for the lowest transverse mode is the uniform mode approximation where we approximate
 the transverse modes by plane waves~\cite{Kreisel09}. This approach is valid if
the thickness $d$ of the YIG film is small compared to the extensions in $y$- and $z$-direction.
Then we find
\begin{eqnarray}
A_\kv &=& h_0 + J S \left[4 - 2 \cos\left(k_y a\right) - 2 \cos\left(k_z a\right)\right] \nonumber\\*
&& - \frac{S}{2} \left(D^{xx}_\kv + D^{yy}_\kv\right) + \frac{\Delta}{3},
\label{eq:Ak}\\
B_\kv &=& - \frac{S}{2} \left(D^{xx}_\kv - D^{yy}_\kv\right),
\end{eqnarray}
where 
 \begin{equation}
 \Delta = \frac{4\pi\mu^2 S }{ a^3}
 \label{eq:Deltadef}
 \end{equation}
is the dipolar energy and the Fourier transformed elements of the dipolar tensor are~\cite{Kreisel09}
\begin{subequations}
 \label{eq:dipdef}
\begin{eqnarray}
D^{xx}_\kv &=& \frac{4 \pi \mu^2}{a^3} \left[\frac{1}{3} - f_\kv\right], \label{eq:Dxx}\\
D^{yy}_\kv &=& \frac{4 \pi \mu^2}{a^3} \left[\frac{1}{3} - \left(1-f_\kv\right) \sin^2 \theta_\kv\right], \label{eq:Dyy}\\
D^{zz}_\kv &=& \frac{4 \pi \mu^2}{a^3} \left[\frac{1}{3} - \left(1-f_\kv\right) \cos^2 \theta_\kv\right], \label{eq:Dzz}\\
D^{yz}_\kv &=& D^{zy}_\kv = -\frac{2\pi\mu^2}{a^3} \sin\left(2\theta_\kv\right),
 \label{eq:Dyz} \\
D^{xy}_\kv &=& D^{yx}_\kv = 0.  \label{eq:Dxy}
\end{eqnarray}
\end{subequations}
The form factor $f_\kv$ is given in  Eq.~\eqref{eq:formdef}. For in-plane wavevectors $D^{yz}_\kv=D^{zy}_\kv$ is the only non-zero off-diagonal matrix element of the dipolar tensor.
Within these approximations the expressions for the magnon energy $\epsilon_\kv$ and the pumping energy $V_\kv$ reduce to Eqs.\eqref{eq:disp_rel} and \eqref{eq:pump_en} of the main text.

\renewcommand{\theequation}{B\arabic{equation}}
\setcounter{subsection}{0}
\setcounter{equation}{0}

\section*{\label{sec:exp_con_corr}
APPENDIX B: EXPANSION IN POWERS OF CONNECTED CORRELATIONS}
In this appendix  we review the method of deriving 
kinetic equations in terms of connected equal-time correlations
developed by J. Fricke in Ref.~[\onlinecite{Fricke97}]. 
In the following we refer to this method as 
the {\it{Fricke approach}}.
In Sec.~\ref{sec:col_int_der} we have used this method to derive the 
leading contributions of the cubic interaction vertices
to the collision integrals appearing in the kinetic equations (\ref{eq:dgl_ci}).
While it is also possible to use the Keldysh formalism\cite{Kamenev11} for this task, the 
Fricke approach is more efficient for our purpose because 
it produces directly a hierarchy of coupled kinetic equations involving only 
equal-time correlations
and provides us with a systematic  decoupling scheme for correlations of arbitrary order.
Note also that the Fricke approach
 generates an expansion of the collision integrals in powers of
connected equal-time correlations and is therefore
very convenient for including  the effect of
time-dependent non-Gaussian correlations in the non-equilibrium dynamics; 
in contrast, 
the Keldysh formalism relies on the
perturbative expansion in terms of single-particle Green functions.

\subsection{Equations of motion}

Consider the bosonic many-body system with second quantized
Hamiltonian $H$ which may explicitly depend on time.
In the Heisenberg picture the time-dependence of an operator $A(t)$ 
is given by the Heisenberg equation of motion,
\begin{equation}
i \frac{\mathrm{d}}{\mathrm{d}t} A(t) = \left[A (t ), H \right] .
\end{equation}
The expectation value of $A( t)$ is given by
\begin{equation}
\langle A \rangle_t = \mathrm{Tr}\left[\rho_0 A(t)\right],
\end{equation}
where the density matrix $\rho_0$ specifies a mixture of states at the initial time  $t_0$. The time-dependence of the expectation value is described by
\begin{equation}
i \frac{\mathrm{d}}{\mathrm{d}t} \langle A \rangle_t = \langle\left[A, H\right]\rangle_t .
\end{equation}
Writing $H = H_0^t + V$, where the one-particle part $H_0^t$ 
contains the terms that are quadratic in the bosonic operators and $V$ describes interactions, we obtain
\begin{equation}
i\frac{d}{dt} \langle A \rangle_t - \langle\left[A, H_0^t\right]\rangle_t = \langle\left[A, V\right]\rangle_t .
\label{eq:eqmA}
\end{equation}
The contribution of the one-particle Hamiltonian $H_0^t$ to the time-evolution of the system is easy to handle.
In order to derive the contribution of the right hand side of Eq.~\eqref{eq:eqmA} containing the interaction Hamiltonian $V$, it is useful to introduce  connected correlations.

\subsection{Connected correlations}

In order to express expectation values of 
an arbitrary set of bosonic operators at the same time
in terms of connected equal-time correlations we introduce the cluster expansion. 
Following again Ref.~[\onlinecite{Fricke97}],
let us consider a set of bosonic operators $B_i$ labeled by  a set of integers $i\in\mathbb{N}$.
The explicit expressions for the connected correlations contain sums over all partitions $P$ of an index set $I$ defined as the set of all non-empty disjoint subsets $J$ of $I$ with $\bigcup_{J \in P} J = I$.  Furthermore, we define 
 $
B_I \equiv B_{i_1} \cdots B_{i_k}
 $
as the product of all operators with indices $i_1,\dots,i_k$, where $i_1<  \ldots
< i_k$ and $I=\{i_1,\dots,i_k\}$.
In our case, the $B_i$ are bosonic field operators, i.e. linear combinations of bosonic creation operators $b_j^\dagger$ and annihilation operators $b_j$.
They obey the commutation relations
\begin{eqnarray}
\big[b_i, b_j\big] &=& 0 , \; \; \; \; 
\big[b_i, b^\dagger_j\big] =  \delta_{i j} .
\label{eq:comm_rel_boson}
\end{eqnarray}
\noindent
Since the operators do not commute in general, we
keep track of their ordering by requireing 
that the indices $i_k$ within the sets $I$ are ordered as denoted above \cite{Fricke97}.

The connected correlations $\langle ... \rangle^c$ can be  defined recursively 
as follows, \cite{Baumann85}
\begin{equation}
\langle B_I \rangle = \sum\limits_{P \in P_I} \prod\limits_{J \in P} \langle B_J \rangle^c ,
\label{eq:def_cc}
\end{equation}
where
$P_I$ refers to the set of all partitions of $I$.
With the help of Eq.~(\ref{eq:def_cc}) we can write $n$-point correlation 
functions as sums over all partitions $P$ of the index set $I$ with each summand being the product of all correlations of the subsets $J \in P$. Note that the correlations preserve the ordering of the indices in the sets $J$.
It is also possible to obtain an explicit expression for the connected correlations,\cite{Fricke97,Baumann85}
\begin{equation}
\langle B_I \rangle^c = \sum\limits_{P \in P_I} \left(-1\right)^{\# P-1} \left(\# P-1\right)! \prod\limits_{J \in P} \langle B_J \rangle ,
\end{equation}
where $\#I$ denotes to the cardinality of the set $I$ which we will refer to as the order of the correlations.
For example, the connected correlations up to third order are,~\cite{Fricke97}
\begin{subequations}
\begin{eqnarray}
\langle B_1 \rangle^c &=& \langle B_1 \rangle , \\
\langle B_1 B_2 \rangle^c &=& \langle B_1 B_2 \rangle - \langle B_1 \rangle \langle B_2 \rangle , \\
\langle B_1 B_2 B_3 \rangle^c &=& \langle B_1 B_2 B_3 \rangle - \langle B_1 B_2 \rangle \langle B_3 \rangle - \langle B_1 \rangle \langle B_2 B_3 \rangle \nonumber\\
&& - \langle B_1 B_3 \rangle \langle B_2 \rangle + 2 \langle B_1 \rangle \langle B_2 \rangle \langle B_3 \rangle .
\end{eqnarray}
\end{subequations}

The commutation relations \eqref{eq:comm_rel_boson} imply 
that correlations with a permuted sequence of field operators differ. Using $\big\langle b_1b_2 \big\rangle^c = \big\langle b_1b_2 \big\rangle - \big\langle b_1 \big\rangle\big\langle b_2 \big\rangle$ it 
follows that the connected one-particle correlations are
\begin{subequations}
\begin{eqnarray}
\langle b_i b_j \rangle^c = \langle b_j b_i \rangle^c, \\
\langle b_i b^\dagger_j \rangle^c = \delta_{i j} + \langle b^\dagger_j b_i \rangle^c .
\end{eqnarray}
\end{subequations}
On the other hand, in correlations of order greater than two the field operators permute 
trivially,\cite{Fricke97}
\begin{subequations}
\begin{eqnarray}
\langle \cdots b_i b_j \cdots \rangle^c &=& \langle \cdots b_j b_i \cdots \rangle^c , \label{eq:cccr1}\\
\langle \cdots b_i b^\dagger_j \cdots \rangle^c &=& \langle \cdots b^\dagger_j b_i \cdots \rangle^c . \label{eq:cccr2}
\end{eqnarray}
\end{subequations}
We note that only connected correlations of order $n = 2$ obey a non-trivial commutation relation and are thus a special case. For this reason, we will refer to one-particle connected correlations as contractions. As we will see later, contractions play an important role for this method.
A proof of Eqs.~\eqref{eq:cccr1} and \eqref{eq:cccr2} can be found in Refs.~[\onlinecite{Baumann85,Fricke97}].

The definition of the cluster expansion can also be extended to fermionic field operators in such a way that we obtain analogous equations. Then  the correlations of order $n \neq 2$ 
anti-commute\cite{Fricke97} and hence sign rules have to be included. 
In this work we are only interested in bosonic operators.

\subsection{Linked-cluster theorem}
We are interested in the time-evolution of $n$-point functions $\langle B_I \rangle_t = \langle B_1 \cdots B_k \rangle_t$, where $B_i$ are linear combinations of bosonic field operators with $i\in I$.
First, we simplify the interaction Hamiltonian $V$ in Eq.~\eqref{eq:eqmA} by assuming the form $V=B_K$. This can be justified by the fact that the equation of motion \eqref{eq:eqmA} is a linear combination of the $B_K$. The linked-cluster theorem discussed  
in this appendix 
still holds for the full interaction Hamiltonian with the form of $V=\sum_K v_K B_K$.

The equation of motion of the expectation value $\langle B_I \rangle$ is given by~\cite{Fricke97}
\begin{equation}
i \frac{\mathrm{d}}{\mathrm{d}t} \langle B_I \rangle_t = \langle\left[B_I, V\right]\rangle_t = \langle B_I B_K - B_K B_I \rangle_t,
\label{eq:eqm_exp}
\end{equation}
where we have chosen $I$ and $K$ to be disjoint without loss of generality.
As the connected correlation obey non-trivial commutation relations in general, we have to keep track of the sequence of indices of the operators inside the expectation values in Eq.~\eqref{eq:eqm_exp}. Therefore we define $I + K$ as the set $I \cup K$ with the order relation given by the order relations of $I$ and $K$ respectively and the condition $i<k\ \forall i \in I,\ k \in K$. Note that the sets $I + K$ and $K + I$ are identical; their order relation differs though.
For $J \in I + K$ we define $\tilde{J}$ as the identical set $J$ but with order relation of 
$K + I$, following Ref.~[\onlinecite{Fricke97}].

It can be shown that there is a linked-cluster-theorem for the equation of motion of the connected correlations which is given by \cite{Fricke97}
\begin{equation}
i \frac{\mathrm{d}}{\mathrm{d}t} \langle B_I \rangle^c_t = \sum\limits_{P \in P_{I,K}^c} \left(\prod\limits_{J \in P} \langle B_J \rangle^c_t - \prod\limits_{\tilde{J} \in P} \langle B_{\tilde{J}} \rangle^c_t \right),
\label{eq:linked_cluster}
\end{equation}
where $P_{I,K}^c$ is the set of all connected diagrams and is defined by
\begin{equation}
P_{I, K}^c \equiv \lbrace P \in P_{I+K} | \forall J \in P : J \cap K \neq \emptyset \rbrace .
\end{equation}
The right-hand side of Eq.~\eqref{eq:linked_cluster} can be further simplified.
We have seen in the above section that only contractions which are one-particle connected correlations obey non-trivial commutation relations. Therefore the correlations $\langle B_J \rangle^c_t$ and $\langle B_{\tilde{J}} \rangle^c_t$ differ only for contraction and are identical for connected correlations of order $n \neq 2$. Obviously, the term within the brackets on the right-hand side of Eq.~\eqref{eq:linked_cluster} is non-zero only if it contains at least one contraction, so that in a diagrammatic representation (see below)
only diagrams that contain contractions of external vertices with the interaction vertex contribute to the equation of motion of connected correlations.

\subsection{Diagrams}
The diagrams introduced here differ from Feynman diagrams because they 
represent differential equations for the correlations. As a consequence, each
diagrams  contain only one interaction vertex. Moeover, each
diagrams describe the time-evolution of a particular correlation at time $t$ so that
there is no time or energy integration involved \cite{Fricke97}. 
Also, we introduce a
 new graphical symbol, the correlation bubble \cite{Schoeller94} representing 
the time-dependent correlations.

\begin{figure}[tb]
\centering
\includegraphics[scale=0.9]{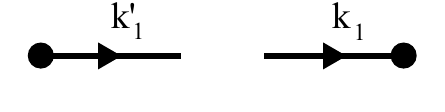}
\caption{External vertices represent annihilation operators with wavevector $\kv'_1$ or creation operators with wavevector $\kv_1$.}
\label{fig:vertices}
\end{figure}

\begin{figure}[tb]
\centering
\includegraphics[scale=0.9]{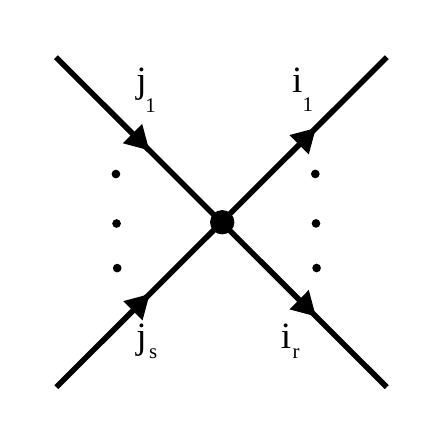}
\caption{Interaction vertices describe the interactions. They are connected with matrix elements $v_{i_1, \dots, i_r; j_1, \dots, j_s}$.}
\label{fig:vertex}
\end{figure}

\begin{figure}[tb]
\centering
\includegraphics[scale=0.9]{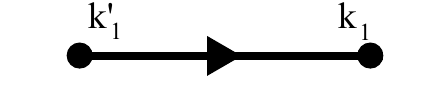}
\caption{The contraction $\big\langle b_{\kv'_1}^\dagger b_{\kv_1} \rangle^c$. Contractions are connected correlations of order two. If the order is larger than two they are correlation bubbles (see Fig.\ref{fig:bubble}).}
\label{fig:contraction}
\end{figure}

\begin{figure}[tb]
\centering
\includegraphics[scale=0.9]{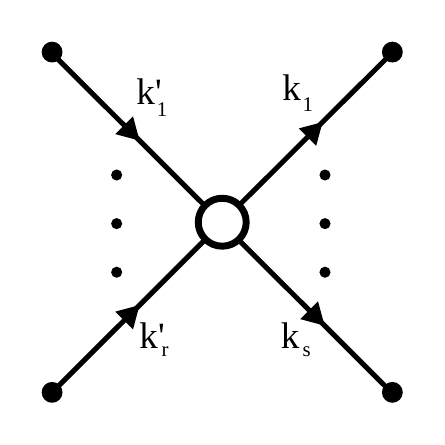}
\caption{Correlation bubbles represent connected correlations, in this case $\langle b_{\kv_1} \cdots b_{\kv_s} b^\dagger_{\kv'_r} \cdots b^\dagger_{\kv'_1} \rangle^c_t$. Note that the order of the connected correlation has to be larger than two. Otherwise it is a contraction (see Fig.\ref{fig:contraction}).}
\label{fig:bubble}
\end{figure}

Let us now  introduce the graphical elements of the diagrams.
External vertices (Fig.~\ref{fig:vertices}) represent annihilation or creation operators. At least one external vertex is contracted with an interaction vertex (Fig.~\ref{fig:vertex}) which represents the interaction associated with a certain matrix element. Contractions (Fig.~\ref{fig:contraction}) are connected one-particle correlations. They are represented by their own graphical element as they play a special role for this method. Connected correlations of order $n \neq 2$ are represented by correlation bubbles (Fig.~\ref{fig:bubble}) \cite{Fricke97}.
As there is not necessarily a conservation of particle numbers for bosons, the number of incoming lines can differ from the number of outgoing lines for interaction vertices and correlation bubbles.


Now we explain the rules for obtaining the time derivative of the $n$ point function $\langle b_{\kv_1} \cdots b_{\kv_s} b^\dagger_{\kv'_r} \cdots b^\dagger_{\kv'_1} \rangle_t$ due to interactions from the diagrams, where $n = r + s$.
The cluster expansion of $\langle [ b_{\kv_1} \cdots b_{\kv_s} b^\dagger_{\kv'_r} \cdots b^\dagger_{\kv'_1}, V ] \rangle_t$ leads to all possible diagrams where vertices are connected with contractions and correlation bubbles.
The resulting diagrams contain $r+s$ external vertices and only one interaction vertex \cite{Fricke97}.
The fact that there is only one interaction vertex simplifies the diagrammatic rules; 
there are no rules regarding time-ordering.

We start with the rule for the prefactor. 
From Eq.~\eqref{eq:eqmA} we get a factor of $(-i)$. Furthermore, we can write the interaction Hamiltonian in the form
\begin{equation}
 V = \frac{1}{r! s!} \sum v_{i_1, \dots, i_r; i'_1, \dots, i'_s} b^\dagger_{i_1} \cdots b^\dagger_{i_s} b_{i'_r} \cdots b_{i'_1} .
\end{equation}
Usually the interaction matrix elements $v_{i_1, \dots, i_r; i'_1, \dots, i'_s}$ fulfill symmetry properties, causing the prefactor of $1/\left(r!s!\right)$ to drop out because permutating annihilation and creation operators in the interaction term gives the same contribution.
However, there exists an exception: if two lines connected to a correlation bubble point into the same direction, permutating the operators yields the same graph and thus the prefactor remains \cite{Fricke97}.

\begin{figure}[tb]
\centering
\includegraphics[scale=0.9]{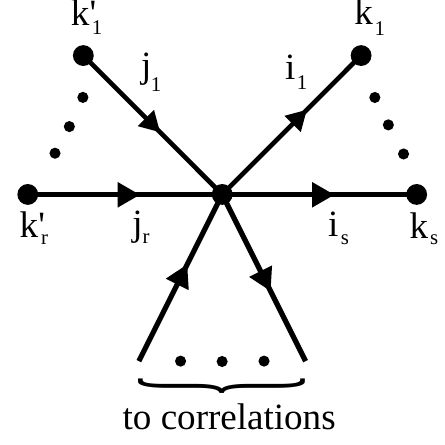}
\caption{Schematic diagram showing the interaction vertex.
In total $r + s$ lines connect the interaction vertex with external vertices. The other lines go to correlation bubbles.}
\label{fig:interaction_vertex}
\end{figure}

The diagrammatic expansion of the
equation of motion for the $n = r+s$-point function has the following structure,
\begin{widetext}
\begin{equation}
\left[\frac{\mathrm{d}}{\mathrm{d}t} + i \left(\epsilon_{\kv_1} + \dots + \epsilon_{\kv_s} - \epsilon_{\kv'_1} - \dots - \epsilon_{\kv'_r}\right)\right] \langle b_{\kv_1} \cdots b_{\kv_s} b^\dagger_{\kv'_r} \cdots b^\dagger_{\kv'_1} \rangle_t = - i \sum_{\mathrm{diagrams}} \frac{1}{2^{n_e}} \sum\limits_{\begin{matrix}i_1, \dots, i_r\\j_1, \dots, j_s\end{matrix}} v_{i_1,\dots,i_r,j_1,\dots,j_s} X_{\mathrm{diagram}} ,
\label{eq:diff_eq_X}
\end{equation}
\end{widetext}
where $X_{\mathrm{diagram}}$ is the collision term containing 
the contractions and correlations and $n_e$ denotes to the number of equivalent pairs of lines.
By comparing Eq.~(\ref{eq:diff_eq_X}) with the general structure \eqref{eq:linked_cluster} 
of the linked cluster expansion
we notice that the collision term contains the difference of the partitions of $\langle b_{\kv_1} \cdots b_{\kv_s} b^\dagger_{\kv'_r} \cdots b^\dagger_{\kv'_1} V \rangle_t$ and the partitions of $\langle V b_{\kv_1} \cdots b_{\kv_s} b^\dagger_{\kv'_r} \cdots b^\dagger_{\kv'_1}\rangle_t$. As connected correlations of order $n \neq 2$ obey trivial commutation relations, there will only be a difference of these two terms due to contractions.
As a result, the collision term is a product of a several factors:
First of all, $X_{\mathrm{diagram}}$ contains all correlations which are denoted by correlation bubbles. Furthermore, there are contributions from contractions. Contractions starting and ending at the interaction vertex give a normal-ordered contribution in the form $-\langle b^\dagger_{\kv_i} b_{\kv_j}\rangle^c_t$, contractions between external vertices give an anti-normal-ordered contribution of the form $\langle b_{\kv_i} b^\dagger_{\kv'_j}\rangle^c_t$.
Finally, there is a contribution from the remaining contractions connecting the external vertices with the interaction vertex. Labeling the diagram as shown in Fig.~\ref{fig:interaction_vertex}, this contribution has the form \cite{Fricke97}
\begin{eqnarray}
\left[\langle b_{\kv_1} b^\dagger_{\bm{i}_1}\rangle^c_t \cdots \langle b_{\kv_s} b^\dagger_{\bm{i}_s}\rangle^c_t \left(-\langle b^\dagger_{\kv'_1} b_{\bm{j}_1}\rangle^c_t\right) \cdots \left(-\langle b^\dagger_{\kv'_r} b_{\bm{j}_r}\rangle^c_t\right) \right.\nonumber\\
\left.- \left(-\langle b^\dagger_{\bm{i}_1} b_{\kv_1} \rangle^c_t\right) \cdots \left(-\langle b^\dagger_{\bm{i}_s} b_{\kv_s} \rangle^c_t\right) \langle b_{\bm{j}_1} b^\dagger_{\kv'_1} \rangle^c_t \cdots \langle b_{\bm{j}_r} b^\dagger_{\kv'_r} \rangle^c_t\right]. \nonumber\\
\end{eqnarray}

Diagrams without at least one contraction connecting an external vertex and an interaction vertex can be omitted, because the contributions of these diagrams vanish due to the fact that only contractions obey non-trivial commutation relations.
Also, we only consider connected diagrams as unconnected diagrams do not contribute to the time-evolution of the correlations according to the linked-cluster theorem \eqref{eq:linked_cluster}. 

\begin{widetext}

\renewcommand{\theequation}{C\arabic{equation}}
\setcounter{subsection}{0}
\setcounter{equation}{0}
\section*{\label{sec:col_int}
APPENDIX C:  COLLISION INTEGRALS FOR PUMPED MAGNONS IN YIG}

Here we use the general formalism outlined in Appendix~B
to derive the  collision integrals due to the cubic interaction vertices in 
the kinetic equations (\ref{eq:dgl_ci}) describing the pumped magnon gas in YIG.
The diagrams contributing to the correlations 
$\langle a^\dagger_\qv a^\dagger_\kqv a_\kv \rangle^c$ and
$\langle a^\dagger_\qv a_\qkv a_\kv \rangle^c$
are shown in Fig.~\ref{fig:diag_2a_1c}. Recall that these diagrams are different from 
Feynman diagrams as they describe the time-evolution of correlations. Therefore they only contain one interaction vertex and there is no time or energy integration associated with the diagrams.
The diagrams shown in  Fig.~\ref{fig:diag_2a_1c} 
represent contributions to the connected three-point correlations
which determine the collision integrals as described in Sec.~\ref{sec:col_int_der},
see Eqs.~(\ref{eq:n_3pc}) and (\ref{eq:p_3pc}).
For the collision integrals associated with 
the diagonal distribution function we obtain for the arrival term
\begin{figure}[tbp]
\centering
\includegraphics[scale=0.3]{fig18a.pdf}
 \hspace{20mm}
\includegraphics[scale=0.3]{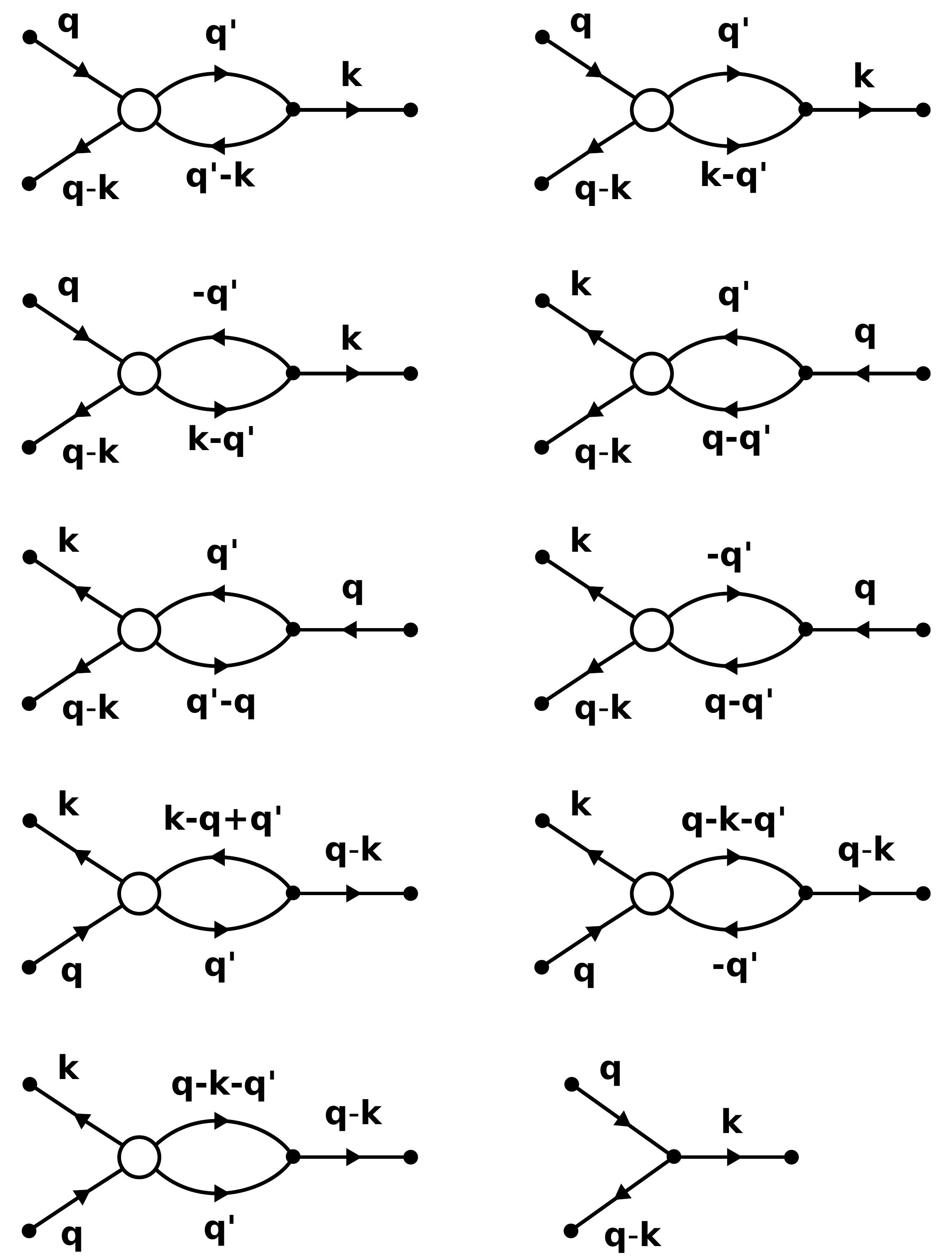}
\caption{Diagrammatic representation of all terms contributing to 
the time-evolution of the correlation $\langle a^\dagger_\qv a^\dagger_\kqv a_\kv \rangle^c$ 
(left set of diagrams)
and to the correlation $\langle a^\dagger_\qv a_\qkv a_\kv \rangle^c$ (right set of diagrams).
}
\label{fig:diag_2a_1c}
\end{figure}

\begin{eqnarray}
 I_{\kv, \mathrm{in}}^n & = & \frac{2 \pi}{N} \sum\limits_\qv \Biggl\{  \frac12 \delta\left(\varepsilon_\kv - \varepsilon_\qv - \varepsilon_\kqv\right) \left|\gam{\kv}{\qv}{\kqv}\right|^2 n^c_\qv n^c_\kqv + \delta\left(\varepsilon_\kv + \varepsilon_\kqv - \varepsilon_\qv\right) \left|\gam{\qv}{\kv}{\qkv}\right|^2 n^c_\qv\left(1+n^c_\qkv\right) \nonumber\\
 & & \hspace{12mm} + \delta  \left( \varepsilon_\kv - \varepsilon_\qv - \varepsilon_\kqv\right) \gam{\kv}{\qv}{\kqv} \sum\limits_{\qv_1',\qv_2'} \delta_{\qv_1'+\qv_2'-\kv,0}  
\biggr[
\gam{\qv_1'}{-\qv_2'}{\kv} e^{-i\omega_0t} \langle\tilde{a}_\qv^\dagger\tilde{a}_\qkv\tilde{a}_{-\qv_2'}^\dagger\tilde{a}_{\qv_1'}\rangle^c 
\nonumber\\
& & \hspace{80mm} + \frac12 \left(\gam{\kv}{\qv_1'}{\qv_2'}\right)^* \langle\tilde{a}_\qv^\dagger\tilde{a}_\qkv\tilde{a}_{\qv_1'}\tilde{a}_{\qv_2'}\rangle^c \biggr]  
\nonumber\\
 & & \hspace{12mm} +\delta\left(\epsilon_\kv+\epsilon_\qkv-\epsilon_\qv\right) \left(\gam{\qv}{\kv}{\qkv}\right)^* \sum\limits_{\qv_1',\qv_2'} 
 \delta_{\qv_1'+\qv_2'-\kv,0}\biggl[ \frac12 \left(\gam{\kv}{\qv_1'}{\qv_2'}\right)^* e^{i\omega_0t} \langle\tilde{a}^\dagger_\qv\tilde{a}^\dagger_{\kqv}\tilde{a}_{\qv_1'}\tilde{a}_{\qv_2'}\rangle^c
 \nonumber\\
& & \hspace{84mm} + \gam{\qv_1'}{\kv}{-\qv_2'} \langle\tilde{a}_\qv^\dagger\tilde{a}_\kqv^\dagger\tilde{a}_{\qv_1'}\tilde{a}_{-\qv_2'}^\dagger\rangle^c \biggr]
 \Biggr\},
 \label{eq:In_in}
\end{eqnarray}
and for the departure term
\begin{eqnarray}
 I_{\kv, \mathrm{out}}^n & = & \frac{2 \pi}{N} \sum\limits_\qv \Biggl\{ \frac12 \delta\left(\varepsilon_\kv - \varepsilon_\qv - \varepsilon_\kqv\right) \left|\gam{\kv}{\qv}{\kqv}\right|^2 n^c_\kv \left(1+n^c_\qv+n^c_\kqv\right) \nonumber\\
 & & \hspace{12mm} + \delta\left(\varepsilon_\kv + \varepsilon_\qkv - \varepsilon_\qv\right) \left|\gam{\qv}{\kv}{\qkv}\right|^2 n^c_\kv \left(n^c_\qkv-n^c_\qv\right)\nonumber\\
 & & \hspace{12mm} + \delta\left(\varepsilon_\kv - \varepsilon_\qv - \varepsilon_\kqv\right) \gam{\kv}{\qv}{\kqv} \sum\limits_{\qv_1',\qv_2'} \delta_{\qv_1'+\qv_2'-\kv,0}\biggl[ \frac12\gam{\qv}{\qv_1'}{\qv_2'} e^{i\omega_0t} \langle\tilde{a}_\kv\tilde{a}_\qkv\tilde{a}^\dagger_{\qv_1'}\tilde{a}^\dagger_{\qv_2'}\rangle^c \nonumber\\
 & & \hspace{28mm} + \left(\gam{\qv_1'}{-\qv_2'}{\qv}\right)^* \langle\tilde{a}_\kv\tilde{a}_\qkv\tilde{a}_{\qv_1'}^\dagger\tilde{a}_{-\qv_2'}\rangle^c - \gam{\qv_2'}{-\qv_1'}{\qkv} e^{i\omega_0t} \langle\tilde{a}_\kv\tilde{a}_\qv^\dagger\tilde{a}_{-\qv_1'}^\dagger\tilde{a}_{\qv_2'}\rangle^c \nonumber\\
 & & \hspace{80mm} - \frac12 \left(\gam{\qkv}{-\qv_1'}{\qv_2'}\right)^* \langle\tilde{a}_\kv\tilde{a}_\qv^\dagger\tilde{a}_{-\qv_1'}\tilde{a}_{\qv_2'}\rangle^c
 \biggr] \nonumber\\
 & & \hspace{12mm} + \delta\left(\varepsilon_\kv + \varepsilon_\qkv - \varepsilon_\qv\right) \left(\gam{\qv}{\kv}{\qkv}\right)^* \sum\limits_{\qv_1',\qv_2'} \delta_{\qv_1'+\qv_2'-\kv,0} \biggl[ \left(\gam{\qv_1'}{\qv_2'}{\qv}\right)^* e^{i\omega_0t} \langle\tilde{a}_\kv\tilde{a}_\kqv^\dagger\tilde{a}_{\qv_1'}^\dagger\tilde{a}_{\qv_2'}\rangle^c \nonumber\\
 & & \hspace{38mm} + \frac12 \gam{\qv}{\qv_1'}{-\qv_2'} \langle\tilde{a}_\kv\tilde{a}_\kqv^\dagger\tilde{a}_{\qv_1'}^\dagger\tilde{a}_{-\qv_2'}^\dagger\rangle^c + \frac12 \gam{\kqv}{\qv_1'}{-\qv_2'} \langle\tilde{a}_\kv\tilde{a}_\qv^\dagger\tilde{a}_{\qv_1'}^\dagger\tilde{a}_{-\qv_2'}^\dagger\rangle^c \nonumber\\
 & & \hspace{28.5mm} + \frac12 \left(\gam{\qv_2'}{\qv_1'}{\kqv}\right)^* e^{i\omega_0t} \langle\tilde{a}_\kv\tilde{a}_\qv^\dagger\tilde{a}_{\qv_1'}\tilde{a}_{\qv_2'}\rangle^c + \left(\gam{\qv_1'}{\qv_2'}{\kqv}\right)^* e^{i\omega_0t} \langle\tilde{a}_\kv\tilde{a}_{\qv}^\dagger\tilde{a}_{\qv_1'}^\dagger\tilde{a}_{\qv_2'}\rangle^c
 \biggr]
 \Biggr\}.
 \label{eq:In_out}
\end{eqnarray}
For the collision integrals of the off-diagonal distribution function we obtain for the arrival term
\begin{eqnarray}
 I_{\kv, \mathrm{in}}^p & = &  \frac{2 \pi}{N} \sum\limits_\qv \Biggl\{ \delta\left(\varepsilon_\kv + \varepsilon_\qkv - \varepsilon_\qv\right) \gam{\qkv}{\qv}{-\kv}\gam{\qv}{\qkv}{\kv} n^c_\qv\left(1+n^c_\qkv\right) \nonumber\\
 & & \hspace{12mm} + \sum\limits_{\qv_1',\qv_2'} \delta\left(\varepsilon_\kv + \varepsilon_\qkv - \varepsilon_\qv\right) \delta_{\kv+\qv_1'-\qv_2',0} \gam{\qkv}{\qv}{-\kv} \biggl[ \gam{\qv_1'}{\qv_2'}{\kv} e^{-i\omega_0t} \langle\tilde{a}_\qv^\dagger\tilde{a}_\qkv\tilde{a}_{\qv_1'}\tilde{a}_{\qv_2'}^\dagger\rangle^c \nonumber\\
 & & \hspace{82mm} + \frac12 \left(\gam{\kv}{\qv_1'}{-\qv_2'}\right)^*  \langle\tilde{a}_\qv^\dagger\tilde{a}_\qkv\tilde{a}_{\qv_1'}\tilde{a}_{-\qv_2'}\rangle^c
 \biggr]\Biggr\},
 \label{eq:Ip_in}
\end{eqnarray}
and for the departure term
\begin{eqnarray}
 I_{\kv, \mathrm{out}}^p & = & \frac{2 \pi}{N} \sum\limits_\qv \Biggl\{ \delta\left(\varepsilon_\kv + \varepsilon_\qkv - \varepsilon_\qv\right) \gam{\qkv}{\qv}{-\kv}\gam{\qv}{\qkv}{\kv} n^c_\kv \left[n^c_\qkv-n^c_\qv\right] \nonumber\\
 & & \hspace{12mm} + \sum\limits_{\qv_1',\qv_2'} \delta\left(\varepsilon_\kv + \varepsilon_\qkv - \varepsilon_\qv\right) \delta_{\kv+\qv_1'-\qv_2',0} \gam{\qkv}{\qv}{-\kv} \biggl[ \frac12 \gam{\qv}{\qv_1'}{\qv_2'} e^{-i\omega_0t} \langle\tilde{a}_\kv\tilde{a}_\qkv\tilde{a}_{\qv_1'}^\dagger\tilde{a}_{\qv_2'}^\dagger\rangle^c \nonumber\\
 & & \hspace{30mm} + \left(\gam{\qv_1'}{-\qv_2'}{\qv}\right)^* \langle\tilde{a}_\kv\tilde{a}_\qkv\tilde{a}_{\qv_1'}^\dagger\tilde{a}_{-\qv_2'}\rangle^c - \gam{\qv_2'}{\qv_1'}{\qkv} e^{-i\omega_0t} \langle\tilde{a}_\qv^\dagger\tilde{a}_\kv\tilde{a}_{\qv_1'}^\dagger\tilde{a}_{\qv_2'}\rangle^c \nonumber\\
 & & \hspace{82mm} - \frac12 \left(\gam{\qkv}{-\qv_1'}{\qv_2'}\right)^* \langle\tilde{a}_\qv^\dagger\tilde{a}_\kv\tilde{a}_{-\qv_1'}\tilde{a}_{\qv_2'}\rangle^c
 \biggr]\Biggr\}.
 \label{eq:Ip_out}
\end{eqnarray}
\end{widetext}


\begin{thebibliography}{99}
\bibitem{footnoteamb}
Eq.~(\ref{eq:n_stat_0}) suggests that there is an ambiguity in the choice of the partition of the connected part $n^c_\kv$ and the contribution $|\vfv{\kv}|^2$ from the expectation values of the magnon operators which is eventually removed by the microscopic collision integrals.
\end{thebibliography}
\end{document}